\documentstyle[emulateapj,psfig]{article}
 
\lefthead{Barger et al.}

\begin{document}
\title{The Nature of the Hard X-ray Background Sources:
Optical, Near-infrared, Submillimeter, and Radio Properties}
\author{A.\,J.\ Barger,$\!$\altaffilmark{1,2,3,4,5}
L.\,L.\ Cowie,$\!$\altaffilmark{1,5}
R.\,F.\ Mushotzky,$\!$\altaffilmark{6}
E.\,A.\ Richards,\altaffilmark{2,7}
}

\altaffiltext{1}{Institute for Astronomy, University of Hawaii,
2680 Woodlawn Drive, Honolulu, Hawaii 96822}
\altaffiltext{2}{Hubble Fellow}
\altaffiltext{3}{Chandra Fellow at Large}
\altaffiltext{4}{Department of Astronomy, University of Wisconsin-Madison,
475 North Charter Street, Madison, WI 53706}
\altaffiltext{5}{Visiting Astronomer, W.\,M.\ Keck Observatory,
jointly operated by the California Institute of Technology and
the University of California}
\altaffiltext{6}{NASA Goddard Space Flight Center, Code 662,
Greenbelt, MD 20771}
\altaffiltext{7}{Arizona State University, 
Tyler Mall PSF-470, Box 871504, Tempe, AZ 85287-1504}

\slugcomment{To appear in the February 2001 Astronomical Journal}

\begin{abstract}

With recent {\it Chandra} observations, at least 75 percent of 
the X-ray background in the $2-10$\ keV energy range is now 
resolved into discrete sources.
Here we present deep optical, near-infrared, submillimeter, and
20\ cm (radio) images, as well as high-quality optical spectra, 
of a complete sample of 20 sources
selected to lie above a $2-10$\ keV flux of
$3.8\times 10^{-15}$\ erg\ cm$^{-2}$\ s$^{-1}$
in a deep {\it Chandra} observation of the Hawaii Deep Survey
Field SSA13. The 13 galaxies with $I<23.5$
have redshifts in the range 0.1 to 2.6. 
Two are quasars, five show AGN signatures, and
six are $z<1.5$ luminous early galaxies whose spectra
show no obvious optical AGN signatures.
The seven spectroscopically unidentified $I>23.5$ sources
have colors that are consistent with evolved early galaxies at 
$z=1.5-3$. 

Only one hard X-ray source is 
significantly detected in an ultradeep submillimeter map;
from the submillimeter to radio flux ratio we estimate a 
millimetric redshift in the range $1.2-2.4$. None of the remaining 
19 hard X-ray sources is individually detected in the 
submillimeter. These results probably reflect the fact that
the 850\ $\mu$m flux limits obtainable with SCUBA are quite close to
the expected fluxes from obscured AGN. 
The ensemble of hard X-ray sources contribute about 10\% of the 
extragalactic background light at submillimeter wavelengths.

From the submillimeter and radio data we obtain
bolometric far-infrared luminosities.
The hard X-ray sources have an average ratio of 
bolometric far-infrared to $2-10$\ keV luminosity of about 60,
similar to that of local obscured AGN. The same ratio for
a sample of submillimeter selected sources is in excess 
of $1100$; this suggests that their far-infrared light
is primarily produced by star formation. 

Our data show that luminous hard X-ray sources are common in 
bulge-dominated optically luminous galaxies with about 10\% 
of the population showing activity at any given time.
We use our measured bolometric corrections with
the $2-10$\ keV extragalactic background light 
to infer the growth of supermassive black holes. Even with a high
radiative efficiency of accretion ($\epsilon=0.1$), the
black hole mass density required to account for the 
observed light is comparable to the local black hole mass density.

\end{abstract}


\section{Introduction}
\label{secintro}

After more than 35 years of intensive work, the origin of the hard 
X-ray background (XRB) is still not fully understood. 
The XRB photon intensity, $P(E)$, with units
photons\ cm$^{-2}$\ s$^{-1}$\ keV$^{-1}$\ sr$^{-1}$,
can be approximated by a power-law, $P(E)=AE^{-\Gamma}$, where
$E$ is the photon energy in keV.
The {\it HEAO1} A-2 experiment (\markcite{marshall80}Marshall et al.\
1980) found that the XRB spectrum from $3-15$\ keV is well described
by a photon index $\Gamma\simeq 1.4$, and this result has been 
confirmed and extended to lower energies by recent analyses of
{\it ASCA} (\markcite{chen97}Chen, Fabian, \& Gendreau 1997;
\markcite{gendreau95}Gendreau et al.\ 1995;
\markcite{miyaji98}Miyaji et al.\ 1998; 
\markcite{ishisaki98}Ishisaki et al.\ 1998)
and {\it BeppoSAX} (\markcite{vecchi99}Vecchi et al.\ 1999) data.
At soft ($0.5-2$\ keV) X-ray energies, 70 to 80 percent of the XRB 
is resolved into discrete sources by the {\it ROSAT} 
satellite (\markcite{hasinger98}Hasinger et al.\ 1998). 
Most of these sources are optically identified 
as unobscured active galactic nuclei (AGN) with spectra that are 
too steep to account for the flat XRB spectrum 
(\markcite{schmidt98}Schmidt et al.\ 1998). 
Thus, an additional population of either absorbed or flat spectrum 
sources is needed to make up the background at higher energies.

XRB synthesis models, constructed within the framework of AGN
unification schemes, were developed to account for the spectral
intensity of the XRB and to explain the X-ray source counts 
in the hard and soft energy bands
(e.g., \markcite{setti89}Setti \& Woltjer 1989;
\markcite{madau94}Madau, Ghisellini, \& Fabian 1994;
\markcite{matt94}Matt \& Fabian 1994;
\markcite{comastri95}Comastri et al.\ 1995;
\markcite{zdziarski}Zdziarski et al.\ 1995;
\markcite{gilli99}Gilli, Risaliti, \& Salvati 1999;
\markcite{wilman99}Wilman \& Fabian 1999;
\markcite{miyaji00}Miyaji, Hasinger, \& Schmidt 2000).
In the unified scheme, the orientation of a molecular torus 
surrounding the nucleus determines the classification of a source.
The models invoke, along with a population of
unobscured type-1 AGN whose emission from the nucleus we see
directly, a substantial population of intrinsically obscured AGN
whose hydrogen column densities of
$N_H\sim 10^{21}-10^{25}$\ cm$^{-2}$
around the nucleus block our line-of-sight. 

A significant consequence of the obscured AGN models is that large
quantities of dust are necessary to cause the obscuration. The
heating of the surrounding gas and dust 
by the nuclear emission from the AGN and
the subsequent re-radiation of this energy into the
rest-frame far-infrared (FIR) suggests that the obscured AGN
should also contribute to the source counts and backgrounds
in the FIR and submillimeter wavelength regimes
(\markcite{almaini99}Almaini, Lawrence, \& Boyle\ 1999;
\markcite{gunn00}Gunn \& Shanks 2000).

Due to instrumental limitations, the resolution of the XRB into 
discrete sources at hard energies had to wait for the arcsecond
imaging quality and high-energy sensitivity of {\it Chandra}.
Deep {\it Chandra} imaging surveys are now detecting sources
in the $2-10$\ keV range that account for
60 to 80 percent of the hard XRB
(\markcite{mushotzky00}Mushotzky et al.\ 2000, hereafter MCBA;
\markcite{giacconi01}Giacconi et al.\ 2001; 
\markcite{garmire01}Garmire et al.\ 2001), 
depending on the XRB normalization.
The mean X-ray spectrum of these sources is in good agreement
with that of the XRB below 10\ keV. Furthermore, because of
the excellent $<1''$ X-ray positional accuracy of
{\it Chandra}, counterparts to the X-ray sources in other
wavebands can be securely identified 
(MCBA; \markcite{brandt00}Brandt et al.\ 2000;
\markcite{giacconi01}Giacconi et al.\ 2001).
In this paper we determine spectroscopic redshifts and 
optical, near-infrared (NIR), submillimeter, and radio properties 
of a complete hard X-ray sample drawn from the MCBA deep
{\it Chandra} observations of the Hawaii Deep Survey Field SSA13.
We take $H_o=65\ h_{65}$\ km\ s$^{-1}$\ Mpc$^{-1}$ and use a 
$\Omega_{\rm M}={1\over 3}$, $\Omega_\Lambda={2\over 3}$ 
cosmology throughout.

\section{Sample and Observations}
\label{secdata}

The present study is based on a 100.9\ ks X-ray map of 
the SSA13 field that was observed with
the ACIS-S instrument on the {\it Chandra} satellite
in December 1999 and presented in MCBA.
The position RA(2000)$=13^h\ 12^m\ 21.4 0^s$,
Dec(2000)$=42^{\circ}\ 41^{'}\ 20.96^{''}$ was placed
at the aim point for the ACIS-S array (chip S3). 
Two energy-dependent images of the back-illuminated S3 chip 
and the front-illuminated S2 chip
were generated in the hard ($2-10$\ keV) and
soft ($0.5-2$\ keV) bands.
MCBA chose the hard band energy range 
to be $2-10$\ keV to facilitate comparisons with {\it ASCA} data.
In the present paper we likewise use the $2-10$\ keV range.
Other recent {\it Chandra} studies of the XRB have used either 
the $2-8$\ keV range (\markcite{brandt00}Brandt et al.\ 2000;
\markcite{horn00}Hornschemeier et al.\ 2000) 
or the $2-7$\ keV range 
(\markcite{fabian00}Fabian et al.\ 2000; 
\markcite{giacconi01}Giacconi et al.\ 2001)
to minimize the backgrounds.

We provide here a detailed description of the data reduction 
techniques that were employed by K.~Arnaud in December 1999
to analyze the SSA13 {\it Chandra} image for the MCBA paper.
Improved X-ray data analysis techniques will be presented
in \markcite{arnaud01}Arnaud et al.\ (2001). 

The X-ray images were prepared with xselect and associated 
ftools at GSFC. ACIS grades 0, 2, 3, 4, and 6 were used, and columns 
at the boundaries of the readout nodes, where event select does
not work properly, were rejected. For the S3 chip the light 
curve was examined and times with high backgrounds were rejected,
giving a total exposure of 95.9\ ks; for the S2 chip the full time 
of 100.9\ ks was used. The images were examined in chip coordinates 
to identify and remove bad columns and pixels. 
For S3 the spectrum of the entire chip was extracted and the Si 
fluorescence line was used to determine the gain.
For the front-illuminated S2 chip radiation damage
caused a systematic change in gain and spectral resolution across 
the chip. An observation from the {\it Chandra} Archive of 
the in-flight calibration sources was analyzed
to determine an approximate correction 
for the spatial gain dependence. The PHA values were
divided by (1-CHIPY*0.0002) to correct the gross gain 
variation. The remaining systematic changes on the fluxes are
small compared to statistical uncertainties. The calibration sources 
were also used to determine the conversion from adjusted PHA to 
energy.

Once the images in the $2-10$\ keV and $0.5-2$\ keV energy bands
were generated for the chips, 
a simple cell detection algorithm was used to find
the sources. The field area was stepped through in $2''$ steps,
and the counts within a $5''$ diameter aperture
(chosen to maximize the enclosed source counts throughout the image
without becoming substantially background dominated)
were measured, together with the background in a
$5''-7.5''$ radius annulus around each position.
The average background was 4.7 counts in a $5''$ cell in the 
$2-10$\ keV S3 image and 1.5 counts in the $2-10$\ keV S2 image.
The distribution of counts is Poisson.
A cut of 17 counts in the $2-10$\ keV S3 image
and 10 counts in the S2 image ensures that there is
less than a 20\% probability of a single spurious
source detection in the entire sample.
The background subtracted counts were searched
for all positions within $4.5'$ of the optical axis 
that satisfied the appropriate count criteria.
The source counts were next corrected 
for the enclosed energy fraction within the $5''$ aperture;
the correction is small for this choice of off-axis radius and was
determined from a second-order polynomial fit to the ratio
of the $5''$ to $10''$ diameter aperture counts as measured 
from the brighter sources in the hard and soft band images.
The final positions were obtained with a centroiding algorithm 
that determined the center of light of the X-ray sources.

The initial source positions were determined from the {\it Chandra}
aspect information, but with a plate scale of $0.4905''$ per pixel.
The absolute pointing was then refined using the 10 
sources with $5\sigma$ radio counterparts (see \S~\ref{secradio}).
The offsets are $3.2''$\ W and $0.3''$\ N with
no roll angle correction. With the offsets applied, 
the dispersion between the radio and X-ray
positions for the 10 sources is $0.4''$. 
The present positions should be more accurate than those given 
in MCBA. Figure~\ref{fig1} shows the {\it Chandra} hard band 
image of SSA13 with the source positions identified by the small 
circles. The large circle illustrates the $4.5'$ radius region 
used in this paper.

The conversion from counts to flux depends on the shape
of the source spectrum. For a power-law spectrum with
a photon index $\Gamma$,
the counts [photons\ s$^{-1}$] in an energy band
$E_1$ to $E_2$ are given by
$N_{E_1-E_2}=\kappa \int_{E_1}^{E_2}\ A(E)\ E^{-\Gamma} dE$,
where $A(E)$ is the effective detector area at energy $E$.
Once the normalization of the spectrum, $\kappa$, is determined
from the observed counts, the flux in the $2-10$\ keV band 
is $f_{HX}=\int_{2}^{10}\ (F(E)\ E^{-1})\ dE$ 
where $F=\kappa\ E^{2-\Gamma}$.
The energy index is $\alpha=\Gamma-1$.

Often $\Gamma=2$ (typical of unabsorbed soft band sources) is 
assumed; however, in the absence of a correction
for the actual opacity, such a procedure underestimates the average
conversion of counts to hard X-ray flux. An alternative approach 
is to determine the value of $\Gamma$ for each 
source from the ratio of soft to hard band counts, 
$N_{0.5-2}/N_{2-10}$. We adopt this approach here; however,
because there is substantial uncertainty in the 
individual source $\Gamma$ values, we use the counts-weighted
mean photon indices of 1.2 (hard) and 1.4 (soft) to determine
the hard and soft band fluxes. Our procedure has the advantage of
giving the correct conversion for the ensemble of sources and 
therefore a correct comparison with the hard XRB, 
but it will result in errors in individual source determinations. 
The typical error is not large; for example, a source with an 
actual $\Gamma=2$ spectrum
would have its $2-10$\ keV flux overestimated by a factor of 1.35.
A subsequent paper (\markcite{arnaud01}Arnaud et al.\ 2001) will give 
the derived fluxes using the best fit spectral parameters to the 
individual sources. 

For the S3 chip, the final flux calibrations were 
made using an array of effective areas versus energy at 12 positions.
For the S2 chip a single conversion factor of
$2.6\times 10^{-11}$\ erg\ cm$^{-2}$\ count$^{-1}$ was used in the 
$2-10$\ keV band and 
$4.5\times 10^{-12}$\ erg\ cm$^{-2}$\ count$^{-1}$
in the $0.5-2$\ keV band.

To check our conversion of counts to flux, we extracted the 
public {\it ASCA} and {\it Chandra} observations of G21.5, a compact 
crab-like supernova remnant with a well-determined simple spectrum. 
The vast difference in angular resolution between {\it ASCA} 
and {\it Chandra} means that we can only make direct source flux 
comparisons for small ($\theta<2'$) or point sources. We can 
also only make direct comparisons of data from instruments with 
different spectral resolutions for a simple spectrum source since 
there the transformation from source counts to flux is less 
model-dependent than it would be for a complex spectrum source.
We derived the column density and power-law index for G21.5 from 
the {\it Chandra} S3 data and found that the
values agreed well with the {\it ASCA} values. The 
measured {\it Chandra} flux was
within 10\% of the {\it ASCA} flux. At present there are no other 
time-stable compact or point sources with simple spectra in the
public database that can be used for such a comparison.

We selected all sources that were more than $15''$ from the 
chip edges and that have $2-10$\ keV fluxes greater than 
$3.8\times 10^{-15}$\ erg\ cm$^{-2}$\ s$^{-1}$, corresponding to 
the above counts cut-offs and the appropriate calibrations.
This detection threshold is slightly higher than the value quoted in
MCBA ($3.2\times 10^{-15}$\ erg\ cm$^{-2}$\ s$^{-1}$) in order to 
obtain a complete flux-limited sample that is uniform over our $4.5'$
radius area on the S2 and S3 chips.
Table~1 details the 20 sources in the resulting 57\ arcmin$^2$ area,
ordered by decreasing hard X-ray flux.
The first six columns in Table~1 include the source identifications
(the MCBA identifications are given in parentheses), 
RA(2000), Dec(2000), $2-10$\ keV flux, $0.5-2$\ keV flux,
and the value of $\Gamma$ required to match
the ratio of the soft to hard X-ray fluxes in the absence of
opacity. The remaining entries in Table~1 are discussed in
subsequent sections.

While the low total X-ray source counts of most of our sources precludes
a detailed analysis on an object-by-object basis, all of our sources
are consistent with the {\it Chandra} PSF. They also all
have rather hard X-ray spectra that are consistent with power-law
or $kT>2$\ keV thermal spectra (\markcite{arnaud01}Arnaud et al.\ 2001).
Thus, the sources are not likely to be emission from groups of galaxies. 
The size constraints on the sources are not sufficiently restrictive
to discriminate between X-ray binaries, hot gaseous atmospheres,
or AGN as the source of the X-ray emission.
We note that an improved comparison between
the expected and observed images shows that object 
CXO J131159.3+123928 (source 1 in the present sample), which was 
thought to be extended by MCBA, is in fact consistent with being a 
point source. 

\subsection{Optical and Near-infrared Imaging}
\label{secimaging}

We used the Low-Resolution Imaging Spectrometer
(LRIS; \markcite{oke95}Oke et al.\ 1995) on the Keck 10\ m 
telescope in March 1997, February 1998 and 1999, and
February and March 2000 to obtain
$B$-band and $I$-band images that cover the {\it Chandra}
SSA13 field.

Wide-field and deep $HK'$ observations
were obtained over a number of runs 
using the University of Hawaii Quick Infrared Camera
(QUIRC; \markcite{hodapp96}Hodapp et al.\ 1996) on the 2.2\ m
University of Hawaii (UH) telescope and the 3.6\ m
Canada-France-Hawaii Telescope.
The $HK'$ ($1.9\ \pm 0.4$\ $\mu$m) filter is described in
\markcite{wainscoat01}Wainscoat \& Cowie (2001). 
The astrometry for the $HK'$ 
image was established by fitting a linear solution to the 75 VLA 
sources (see \S~\ref{secradio}) with bright counterparts in
the NIR image. The solution has a plate scale of $0.1891''$
per pixel and a rotation of $-0.27$\ deg from a 
standard N-E orientation. The dispersion of the
radio and NIR positions is $0.43''$. Using a large overlap sample
of sources in the optical and NIR images, a third-order polynomial 
fit to the distortion in each LRIS $I$ and $B$ image was determined.
Using the 90 VLA sources with bright $I$ counterparts in the 
corrected image, we find a dispersion of $0.67''$ between the radio 
and optical positions.

The excellent $<1''$ X-ray positional accuracy permits the
secure identification of the optical counterparts to the
X-ray sources. Figure~\ref{fig2} shows thumbnail $B$-band
images of all 20 hard X-ray sources listed in Table~1.
(Thumbnail $I$-band images can be found in MCBA.)
In selecting the optical counterparts, we considered
only $I\le 24.5$ sources within a $1.5''$ radius of the nominal 
X-ray position. The optical separations are given in column 10 of 
Table~1. Sixteen of the sources have one such counterpart, and 
none has more than one. None of the offsets exceed $1''$, and 
the dispersion of the offsets 
is $0.5''$. Monte Carlo simulations with a randomized sample show that
the average number of spurious identifications with an optical
counterpart within $1''$ is 0.5. At 95 percent confidence,
less than 2 of the 16 identifications are spurious.
For the 16 sources the magnitudes were measured at the
optical center; for the remaining sources, the magnitudes
were measured at the nominal X-ray position. For most of the sources
the magnitudes were measured in $3''$ diameter apertures and 
corrected to approximate total magnitudes using an average offset 
(\markcite{cowie94}Cowie et al.\ 1994); henceforth, we refer to these
as corrected $3''$ diameter magnitudes. However,
for the bright extended sources (7, 18, 19)
we used $20''$ diameter aperture magnitudes and applied no correction;
these magnitudes may be as much as a magnitude brighter than the
corrected $3''$ diameter magnitudes given in MCBA.
The $B$, $I$, and $HK'$ magnitudes are given in columns 
7, 8, and 9 of Table~1. 
The $1\sigma$ limits are approximately $B=27.6$ and $I=25.9$.
The $1\sigma$ limits for the $HK'$ magnitudes are not uniform over 
the field and are given individually in parentheses after the 
$HK'$ magnitudes in Table~1.

\subsection{Keck Spectroscopy}
\label{seckeck}

We obtained high quality optical spectra for 19 of the 
20 hard X-ray sources using LRIS slit-masks
on the Keck 10\ m in March and April 2000. Source 14 was
not observed because of mask design constraints. 
For the sources with $I\le 24.5$ counterparts, we positioned the slit
at the optical center.
For the remaining sources, we positioned the slit at the 
X-ray centroid position.
We used $1.4''$ wide slits and the 300 lines\ mm$^{-1}$
grating blazed at 5000\,\AA, which gives a
wavelength resolution of $\sim 16$\,\AA\ and a wavelength coverage of
$\sim 5000$\,\AA. The wavelength range for each object depends on the
exact location of the slit in the mask but is generally between
$\sim5000$ and 10000\,\AA. 
The observations were 1.5\ hr per slit mask,
broken into three sets of 0.5\ hr exposures.
Fainter objects were observed a number of times;
the longest exposure was 6\ hrs.
Conditions were photometric with seeing $\sim 0.6''-0.7''$ FWHM.
The objects were stepped along the slit by $2''$ in each
direction, and the sky backgrounds were removed using the median of
the images to avoid the difficult and time-consuming problems of
flat-fielding LRIS data. Details of the spectroscopic reduction
procedures can be found in \markcite{cowie96}Cowie et al.\ (1996).

We successfully obtained redshift identifications for all 13 sources
brighter than $I=23.5$\ mag;
the spectra are shown in Fig.~\ref{fig3},
and the redshifts are given in column 11 of Table~1.
We classify the spectra into
three general categories: (i)\ quasars (broad-line sources),
(ii)\ AGN (narrow and weak-line sources), and (iii)\ optically 
`normal' galaxies (no AGN signatures in the optical).
Henceforth, we denote these categories by {\it q}, {\it a}, and 
{\it n}, respectively. We also denote spectroscopically unidentified 
sources by {\it u} and our one source with a millimetric redshift 
(see \S~\ref{secdetection}) by {\it m}.

Two sources (sources 3 and 6) are the quasars previously
known to be in the field (\markcite{windhorst}Windhorst et al.\ 1995;
\markcite{campos}Campos et al.\ 1999).
Their spectra are very similar, and both coincidentally lie 
at $z=2.565$. These quasars are radio quiet.

Five sources (1, 2, 5, 11, and 15) show emission line characteristics
that may be indicative of AGN activity. Sources 1 and 5
show [O\,II], Ne\,III], and weak Ne\,V] 
emission, along with Ca H and K and G-band absorption.
Sources 11 and 15 show narrow Ly$\alpha$ and CIV emission.
Source 2 shows a P-Cygni profile in Mg\,II and broad absorption in Fe\,II.

Six sources (4, 7, 10, 12, 18, and 19)
show no indication of an active nucleus in their optical spectra.
We call these `normal' galaxies since they have absorption 
and emission line properties which are common in optically selected
field samples.
Source 4 shows H$\alpha$ and weak [O\,II] and [O\,III] emission
and H$\beta$ absorption. 
Source 7 shows H$\alpha$, [O\,II], and [O\,III] emission
and H$\beta$ absorption. 
Source 10 shows narrow [O\,II] emission and Mg\,II absorption. 
Source 12 has weak [O\,II] and H$\beta$ emission and strong
[O\,III] absorption. There may be hints of NeIII] and NeV].
Source 18 is rather unusual in that it has no H$\alpha$ 
while N\,II and S\,II
are in emission. A high N\,II/H$\alpha$ ratio has been used 
to classify objects as AGN (\markcite{keel}Keel et al.\ 1985), 
but the absence of H$\alpha$ in source 18 is difficult to 
understand: in \markcite{veilleux87}Veilleux \& Osterbrock (1987)
the highest ratio of NII to H$\alpha$ is 3:1, and no photoionization 
models (\markcite{ferland83}Ferland \& Netzer 1983) have ratios 
larger than 2:1. Source 19 has H$\alpha$ emission but otherwise 
only absorption features.

Optically `normal' X-ray luminous galaxies
were thought to be relatively rare, unusual objects, perhaps 
explained by beaming (\markcite{elvis81}Elvis et al.\ 1981;
\markcite{moran96}Moran et al.\ 1996;
\markcite{tananbaum97}Tananbaum et al.\ 1997).
They are hard to find by association since
small X-ray error boxes are required to be certain of their
identification (e.g., discussion in
\markcite{schmidt98}Schmidt et al.\ 1998). The very large
surface density of such sources in our sample, $\sim400$\ deg$^{-2}$,
indicates that they are common. In fact, in our sample they are
much more common than quasars.
There are two plausible explanations for the lack of observed 
optical AGN characteristics: i)\ absorption due to dust and gas 
or ii)\ an actual lack of ultraviolet/optical emission, as is the case 
in many low luminosity objects (\markcite{ho99}Ho et al.\ 1999). 
The line of sight column densities inferred from the X-ray spectra
in \S~\ref{secoptdepths} are sufficiently large to 
obscure the optical AGN signatures, but the extent
will depend on the geometry. 

\subsection{Submillimeter Observations}
\label{secsmm}

The submillimeter observations were made with the SCUBA
instrument (\markcite{holland99}Holland et al.\ 1999)
on the James Clerk Maxwell Telescope. 
SCUBA jiggle map observations were taken in mostly excellent
observing conditions during runs in 
February 1999 (7 observing shifts),
February 2000 (0.5 shift), and May-June 2000 (3.5 shifts).
The maps were dithered to prevent any regions of the sky from 
repeatedly falling on bad bolometers. The chop throw was
fixed at a position angle of 90\ deg so that the negative beams 
would appear $45''$ on either side east-west of the positive beam.
Regular ``skydips'' (\markcite{manual}Lightfoot et al.\ 1998) 
were obtained to measure the zenith atmospheric opacities at 
450 and 850\ $\mu$m, and the 225\ GHz sky
opacity was monitored at all times to check for sky stability.
The median 850\ $\mu$m optical depth for all nights together 
was 0.185. Pointing checks were performed every 
hour during the observations on the blazars 1308+326 or cit6. 
The data were calibrated using $30''$ diameter aperture 
measurements of the positive
beam in beam maps of the primary calibration source Mars and the
secondary calibration sources OH231.8, IRC+10216, and 16293-2422.

The data were reduced in a standard and consistent way using the
dedicated SCUBA User Reduction Facility
(SURF; \markcite{surf}Jenness \& Lightfoot 1998).
Due to the variation in the density of bolometer samples across 
the maps, there is a rapid increase in the noise levels at the 
very edges. The low exposure edges were clipped from our images.

The SURF reduction routines arbitrarily normalize all the data
maps in a reduction sequence to the central pixel of the first
map; thus, the noise levels in a combined image are
determined relative to the quality of the central pixel in the
first map. In order to determine the absolute noise levels of
our maps, we first eliminated the $\gtrsim 3\sigma$ real sources in
each field by subtracting an appropriately normalized version
of the beam profile. We then iteratively adjusted the noise
normalization until the dispersion of the signal-to-noise
ratio measured at random positions became $\sim 1$.
Our noise estimate includes both fainter sources and correlated noise.

We centered on the positions of the hard X-ray sources and 
measured the submillimeter fluxes 
using beam-weighted extraction routines that include
both the positive and negative portions of the chopped images,
thereby increasing the effective exposure times.
The 850\ $\mu$m submillimeter fluxes and $1\sigma$ uncertainties are 
summarized in column 12 of Table~1. For most of the objects the 
$1\sigma$ level is in the $1-2.5$\ mJy range; however, for the two 
sources in the region where there is an ultradeep 
SCUBA image (\markcite{barger98}Barger et al.\ 1998) the
$1\sigma$ detection threshold is $0.6-0.7$\ mJy.

\subsection{Radio Observations}
\label{secradio}

A very deep 1.4\ GHz VLA radio map of the
SSA13 region was obtained by 
\markcite{richards01}Richards et al.\ (2001)
using an 100 hr exposure in the A-array configuration.
The primary image covers a $40'$ diameter region with an effective
resolution of $1.6''$ and a $5\sigma$ limit of
25\ micro-Jansky ($\mu$Jy). The radio fluxes
were measured in $2.4''$ boxes centered
on the X-ray positions; 
these fluxes are given in column 13 of Table~1.
Of the 20 hard X-ray sources, 16 are detected in the radio above
a $3\sigma$ threshold of 15 $\mu$Jy, including 10 of the 13
sources with spectroscopic redshifts.
The radio-X-ray offsets are given in column 14 for the 10
sources with $5\sigma$ radio detections within $1.5''$ of the X-ray
source. The absolute radio positions are known to 
$0.1-0.2''$\ {\it rms}. The dispersion between the radio and 
X-ray positions is $0.4''$, and
the maximum separation is $0.7''$.
The radio to optical ratios for the hard X-ray sources
are consistent with the sources being
radio quiet. The radio properties  will be discussed in more detail in
\markcite{richards01}Richards et al.\ (2001).

\section{X-ray Properties of the Hard X-ray Sample}
\label{secxray}

\subsection{Optical Depths}
\label{secoptdepths}

\begin{figure*}[tb]
\centerline{\psfig{figure=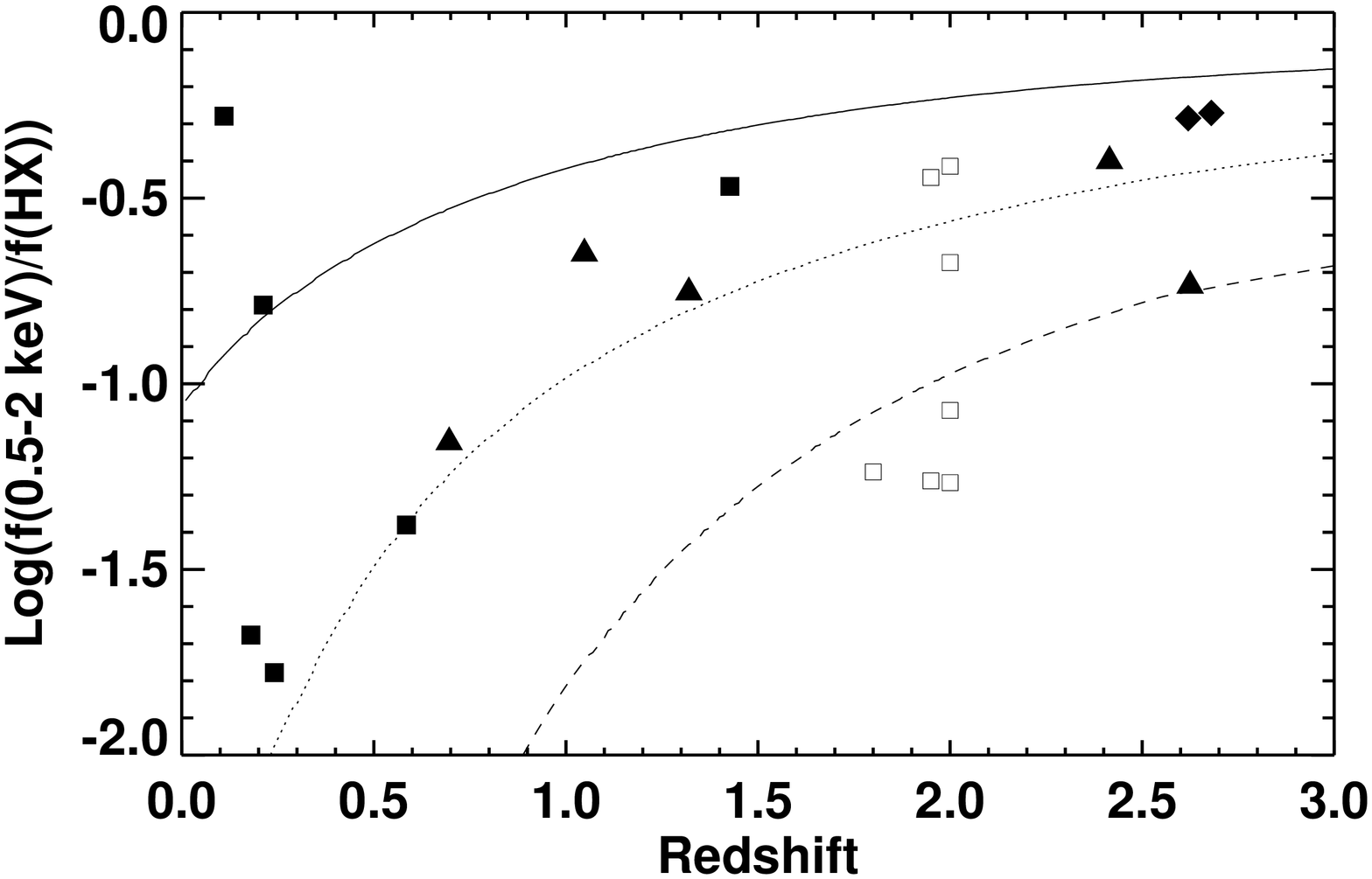,width=3.5in,angle=90}}
\figurenum{4}
\figcaption[]{
Ratio of the soft to hard X-ray fluxes versus redshift. Quasars are
denoted by filled diamonds, AGN by filled triangles, `normal'
galaxies by filled squares, and spectroscopically unidentified
sources by open squares. The latter, excluding source 8,
are placed at $z=2$; source 8 is placed at its millimetric
redshift of 1.8 (see \S~\ref{secdetection}).
The two quasars have been slightly offset in redshift for clarity,
as have the unidentified sources 16 and 17.
As illustrated in the figure, the observed flux ratios can be
described by an input $\Gamma=2$ spectrum and
a narrow range of $N_H$;
curves are for $N_H=2\times 10^{22}$\ cm$^{-2}$ (solid),
$N_H=10^{23}$\ cm$^{-2}$ (dotted),
and $N_H=3\times 10^{23}$\ cm$^{-2}$ (dashed).
\label{fig4}
}
\end{figure*}

With the exception of two `normal' galaxies with $\Gamma<-1.5$,
the photon indices given in Table~1 range from $-0.4$ to $1.82$.
The two quasars have $\Gamma=1.75$ and $\Gamma=1.80$,
consistent with most unabsorbed soft band sources, 
whereas most of the remaining sources have photon indices that
suggest substantial line-of-sight optical depths.

If we generate counts-weighted mean photon indices 
for each population separately, we find 0.8 for the `normal' 
galaxies, 0.9 for the AGN, 1.8 for the two quasars, and 
1.2 for the spectroscopically unidentified sources. 
The `normal' galaxies are 
presumably those where the AGN are the most highly obscured.
Since the effective column density, $N_{eff}$, for 
observed-frame absorption in a source is related to
the true hydrogen column density, $N_H$, by
$N_{eff}\sim N_H/(1+z)^{2.6}$, flux corrections for absorption
effects become less important with increasing redshift.
The spectroscopically unidentified sources
are likely higher redshift analogs of the `normal' galaxies 
(see \S~\ref{seccolors}) but
have softer spectra because of this redshift dependence of the
absorption.

In Fig.~\ref{fig4} we plot the logarithm of the ratio of the soft
to hard X-ray fluxes versus redshift.
Throughout the paper we use the notation of filled diamonds
for quasars, filled triangles for galaxies with AGN 
signatures, filled squares for galaxies with apparently `normal' 
spectra, and open squares for spectroscopically 
unidentified sources. We overlay on the data fixed $N_H$ curves
which we generated assuming an intrinsic $\Gamma=2$ power-law 
spectrum and photoelectic cross sections
computed for solar abundances by 
\markcite{morrison83}Morrison \& McCammon (1983).
Over the energy range $0.5-7$\ keV, which determines the
correction for these column densities and redshifts, 
the cross-section,
$\sigma(E)$, can be well-approximated by a single power-law 

$$\sigma(E)=2.4\times 10^{-22}\ E^{-2.6}\ {\rm cm}^{-2}$$

\noindent
with $E$ in keV. For all but one of the spectroscopically 
identified galaxies the X-ray flux ratios can be described
by a rather narrow range of neutral hydrogen column densities from
$N_H=2\times 10^{22}$\ cm$^{-2}$ to $3\times 10^{23}$\ cm$^{-2}$,
although more sophisticated models with scattering could permit
higher opacities. The $N_H$ values and the true power-law indices
are best determined directly from the X-ray spectra, as shall be
discussed in a subsequent paper 
(\markcite{arnaud01}Arnaud et al.\ 2001).

Could the absence of sources with column densities above 
$N_H=3\times 10^{23}$\ cm$^{-2}$ be a selection effect?
This is possible at low redshifts since the ratio of the 
absorbed to the actual $2-10$\ keV flux drops rapidly 
for $N_{eff}>10^{23}$\ cm$^{-2}$;
however, at high redshifts this $N_{eff}$ corresponds to larger
values of $N_H$ than are observed (see Fig.~\ref{fig4}). 
Even if we place all of the spectroscopically unidentified sources
at $z\gg 1$, at most 3 of the 14 objects with $z>1$
have column densities above $N_H=3\times 10^{23}$\ cm$^{-2}$.
%This suggests that we are in fact seeing most
%of the obscured AGN in the present sample.
The simplest interpretation of Fig.~\ref{fig4} 
is that we are seeing most of the obscured AGN in 
the present sample. However, Compton-thick sources might
be missed completely from the $2-10$\ keV sample, and 
these sources may be needed 
to explain the 30\ keV peak in the XRB. Unfortunately, this issue
cannot be decided until either more precise information is obtained on
the energy distribution of the individual sources, which, in
principle, could deviate from power-law behavior, or until
we can analyze the hardness of even fainter $2-10$\ keV sources.

\subsection{Redshift Distribution}

\begin{figure*}[tb]
\centerline{\psfig{figure=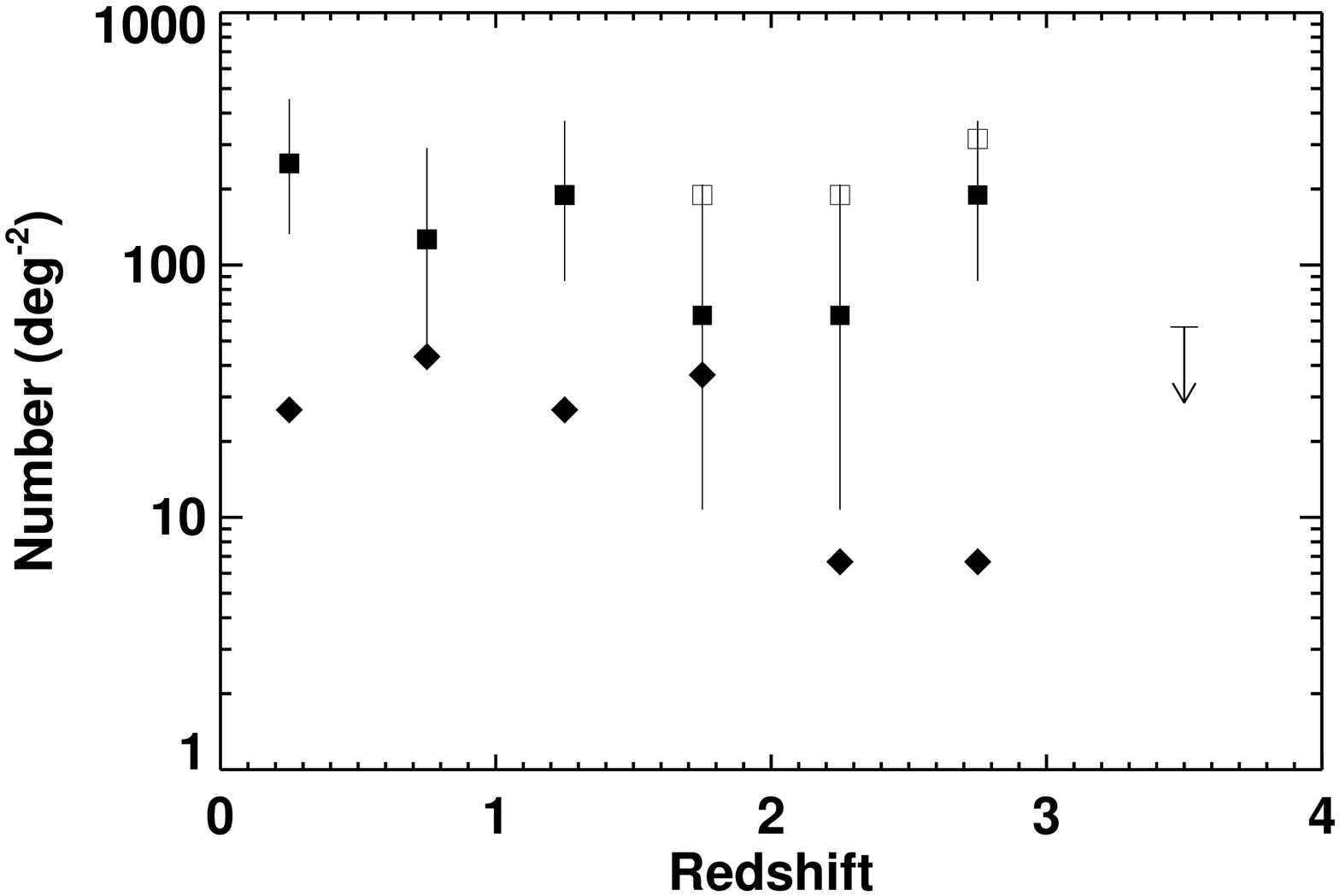,width=3.5in,angle=90}}
\figurenum{5}
\figcaption[]{
Filled squares with $1\sigma$ uncertainties show the surface
density of the spectroscopically identified hard X-ray
sources in $\Delta z=0.5$ redshift bins. For $z=3-4$ we show the
1$\sigma$ upper limit as the downward pointing arrow. Open squares
include the unidentified sources spread uniformly from $z=1.5$ to
$z=3$. Filled diamonds show the surface density of
the deep {\it ROSAT} soft X-ray survey
of the Lockman Hole (Schmidt et al.\ 1998).
\label{fig5}
}
\end{figure*}

The redshift distribution of the spectroscopically
identified sample is shown in Fig.~\ref{fig5}, where we
we plot the surface density of the sources
in $\Delta z=0.5$ redshift bins (filled squares) with $1\sigma$
uncertainties. Below $z=3$ the data are consistent
with a constant surface density. If we also include
the unidentified sources spread uniformly through
the $z=1.5$ to $z=3$ redshift interval (open squares),
then the constancy with redshift becomes even more evident.
The surface density is $\sim 200$ sources per square degree
per bin or $\sim 400$ sources per square degree per unit redshift.

The present data sample is too small to justify a
detailed examination of the luminosity function. However,
the redshift distribution of the hard selected
sample is very similar to that of previous soft selected samples
with similar limiting sensitivities. We illustrate this
in Fig.~\ref{fig5} by overplotting the redshift distribution
of the Lockman Hole {\it ROSAT} sample of 
\markcite{schmidt98}Schmidt et al.\ (1998)
(filled diamonds). The limiting flux of this sample
is $5.5\times 10^{-15}$\ erg\ cm$^{-2}$\ s$^{-1}$ ($0.5-2$\ keV) 
which, for a source with $\Gamma=2$, would correspond to a limiting
flux of $4.7\times 10^{-15}$\ erg\ cm$^{-2}$\ s$^{-1}$
($2-10$\ keV), similar to our limiting
flux of $3.8\times 10^{-15}$\ erg\ cm$^{-2}$\ s$^{-1}$
($2-10$\ keV).
The absolute surface density of the soft sample
is far below that of the hard sample, indicating that
most of the hard X-ray sources are substantially
obscured; however, the redshift distributions are rather
similar. The luminosity function evolution of soft
X-ray samples has been extensively analyzed and shows
a rapid rise between $z=0$ and $z=1.5$, followed by relative
constancy at higher redshifts 
(\markcite{miyaji00}Miyaji et al.\ 2000). 
While a proper analysis of the evolution of the $2-10$\ keV 
luminosity function must await larger samples and a better 
understanding of the optical depths and K-corrections, the first 
impression is that the behavior will be similar to that inferred
from the soft samples but with a much higher normalization
of the luminosity function (about 7 times higher).

\subsection{Luminosities}
\label{seclum}

The intrinsic flux, $F_{int}$, is related to the observed flux, 
$F_{obs}$, by

$$F_{int}=F_{obs}\ (1+z)^{2-\Gamma}$$

\noindent
For an unabsorbed spectrum with $\Gamma=2$ the K-correction
vanishes. An unabsorbed spectrum may be appropriate
at the higher energies where opacity effects are not important.
However, we believe that we can obtain a slightly improved
estimate for the $2-10$\ keV luminosities by allowing for the
average effects of the opacity as follows. 
In calculating our hard X-ray luminosities, we normalized the flux 
at 4\ keV for the $\Gamma=2$ spectrum to the flux at 4\ keV 
calculated over the $2-10$\ keV energy range for a spectrum with
counts-weighted mean photon index $\Gamma=1.2$ (see \S~\ref{secdata}).
Then

$$L_{HX} = 4\pi\ d_L^2\ (0.85)\ f_{HX}$$

\noindent
where $f_{HX}$ is the $2-10$\ keV flux of Table~1
computed with the same $\Gamma=1.2$ assumption.
Our hard X-ray luminosities are given in column~3
of Table~\ref{tab2}. If we had instead used $\Gamma=2$,
the computed $2-10$\ keV fluxes would be lower
by a factor of 1.35 and our X-ray luminosities would be
lower by a factor of 1.15. The X-ray luminosities of
\markcite{fabian00}Fabian et al.\ (2000) and
\markcite{horn00}Hornschemeier et al.\ (2000) were based
on $\Gamma=2$; in later comparisons with their luminosities, 
we ignore this difference since it is small compared to
other uncertainties.

\begin{figure*}[tb]
\centerline{\psfig{figure=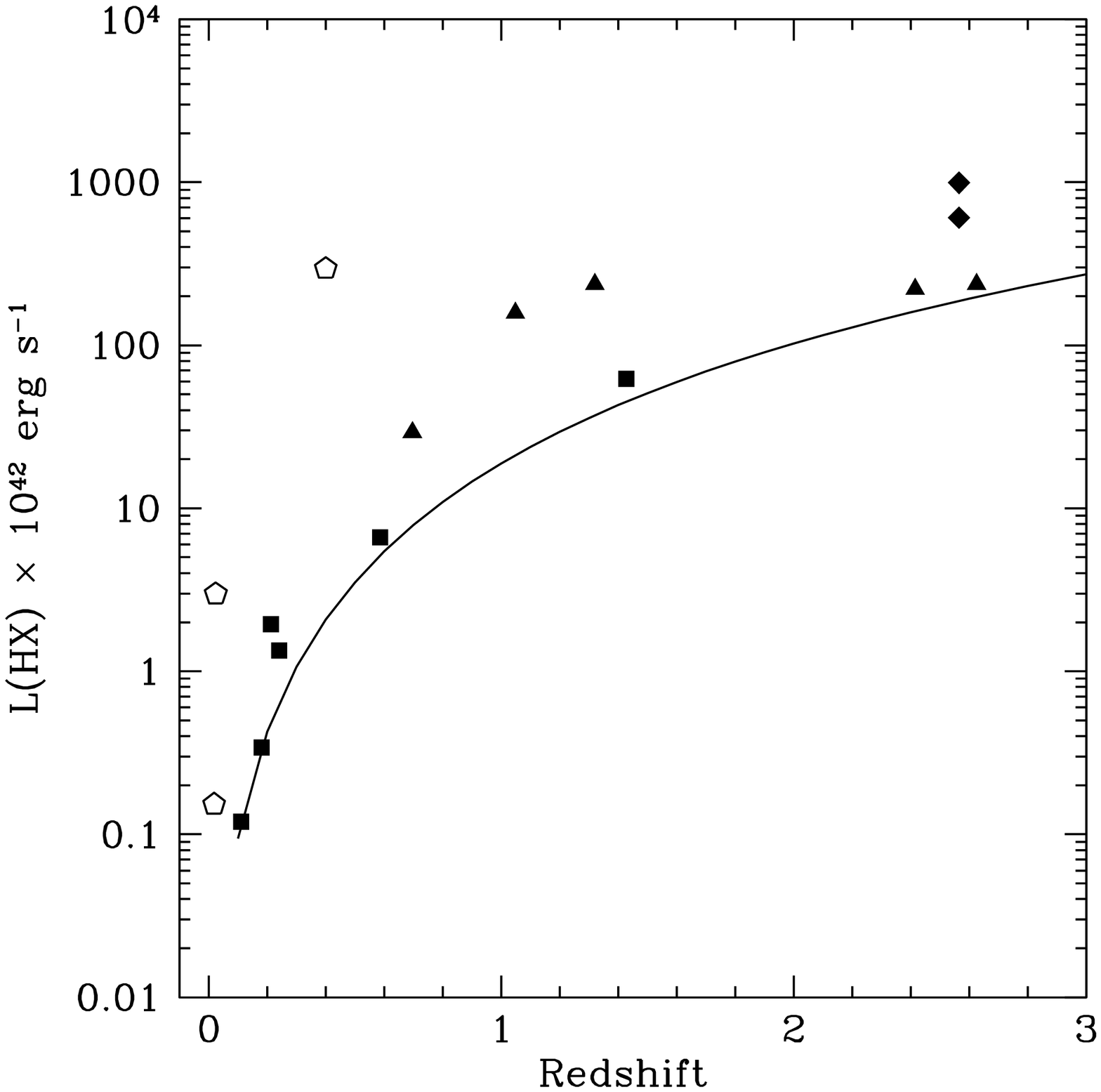,width=3.0in,height=3.0in}}
\figurenum{6}
\figcaption[]{
Hard X-ray luminosities versus redshift for the thirteen $I<23.5$
hard X-ray sources with spectroscopic identifications.
Quasars are denoted by filled diamonds, AGN by filled triangles,
and `normal' galaxies by filled squares. The solid curve
shows the detection limit.
The open pentagons represent Arp~220 (least luminous),
NGC~6240, and PG~1543+489 (most luminous).
\label{fig6}
}
\end{figure*}

Figure~\ref{fig6} shows our hard X-ray luminosities versus redshift.
Here the solid curve represents the detection limit.
The open pentagons denote the ultraluminous starburst galaxy
Arp~220, the ultraluminous, highly obscured AGN NGC~6240, and 
the radio quiet quasar PG~1543+489, in ascending $L_{HX}$ order
(see \S~\ref{secsmmradio}) for comparison.
The hard X-ray luminosities of the spectroscopically identified
sources range from just over $10^{41}$\ erg\ s$^{-1}$
to $\sim 10^{45}$\ erg\ s$^{-1}$. 
Sources at low redshift ($z<1$) do not have the high X-ray
luminosities of the sources at high redshift. To determine whether
there are also low luminosity sources at high redshift will require
{\it Chandra} observations that probe to much deeper flux levels.

Even though the `normal' galaxies,
which typically fall near the detection threshold,
are systematically less luminous than the AGN and quasars,
they are still extremely X-ray luminous;
only the two lowest luminosity sources even overlap
the local galaxy populations (\markcite{fabbiano}Fabbiano 1989).
Furthermore, the `normal' galaxies are generally more luminous
than the hard X-ray
sources in the nuclei of nearby giant elliptical galaxies
in the Virgo and Fornax clusters that were chosen for study on the
basis of their black hole properties; these have $1-10$\ keV
luminosities of $2\times 10^{40}$\ erg\ s$^{-1}$
to $2\times 10^{42}$\ erg\ s$^{-1}$
(\markcite{allen00}Allen, Di Matteo, \& Fabian\ 2000).

\begin{deluxetable}{ccrrrrr}
\renewcommand\baselinestretch{1.0}
\tablewidth{0pt}
\parskip=0.2cm
\tablenum{2}
\small
\tablehead{
Number & $z$ & $L_{HX}$ &
$L_{OPT}$ & $L^{radio}_{FIR}$ & $L^{submm}_{FIR}$ & $L_{BOL}$ \cr
& & ($10^{42}$\ ergs\ s$^{-1}$)
& ($10^{42}$\ ergs\ s$^{-1}$)
& ($10^{42}$\ ergs\ s$^{-1}$) & ($10^{42}$\ ergs\ s$^{-1}$)
& ($10^{42}$\ ergs\ s$^{-1}$) \cr
}
\startdata
0 & 2.000\tablenotemark{u} & 1000 & 480 & 4500 & $<8800$ & 9400 \cr
1 & 1.048\tablenotemark{a} & 150 & 41 & 3600 & $<19000$ & 4300 \cr
2 & 1.320\tablenotemark{a} & 230 & 2400 & $<1200$ & $<17000$ & $<4600$ \cr
3 & 2.565\tablenotemark{q} & 930 & $2.8\times 10^5$ & 28000 & $<51000$
& $3.1\times 10^5$\cr
4 & 0.212\tablenotemark{n} & 1.9 & 4.6 & $<18$ & $<1800$ & $<31$ \cr
5 & 0.696\tablenotemark{a} & 28 & 140 & 800 & $<7900$ & 1000 \cr
6 & 2.565\tablenotemark{q} & 570 & 49000 & 11000 & $<41000$ &
$62000$ \cr
7 & 0.241\tablenotemark{n} & 1.3 & 8.0 & 130 & $<2100$ & 140 \cr
8 & 1.800\tablenotemark{m} & 120 & 320 & 6500 & 5200 & 7400 \cr
9 & 2.000\tablenotemark{u} & 130 & 270 & 4700 & $<14000$ & 5500 \cr
10 & 1.427\tablenotemark{n} & 59 & 520 & 1800 & $<12000$ & 2500 \cr
11 & 2.415\tablenotemark{a} & 210 & 4400 & 8200 & $<16000$ & 14000 \cr
12 & 0.585\tablenotemark{n} & 6.5 & 160 & 220 & $<5100$ & 400 \cr
13 & 2.000\tablenotemark{u} & 130 & 94 & $<3200$ & $<14000$ & $<3800$ \cr
14 & 2.000\tablenotemark{u} & 120 & 140 & 6600 & $<12000$ & 7300 \cr
15 & 2.625\tablenotemark{a} & 220 & 2200 & $<5900$ & $<10000$ & $<9000$ \cr
16 & 2.000\tablenotemark{u} & 110 & 110 & 3600 & $<5600$ & 4200 \cr
17 & 2.000\tablenotemark{u} & 110 & 650 & 16000 & $<9600$ & 18000 \cr
18 & 0.110\tablenotemark{n} & 0.12 & 23 & 27 & $<680$ & 51 \cr
19 & 0.180\tablenotemark{n} & 0.34 & 14 & 77 & $<1300$ & 93
\enddata
\label{tab2}
\end{deluxetable}

\subsection{Contribution to the Hard XRB}
\label{secxrb}

The integrated $2-10$\ keV light of the 20 hard X-ray sources 
in the sample corresponds to an extragalactic background light 
(EBL) of $1.34\times10^{-11}$\ erg\ cm$^{-2}$\ s$^{-1}$\ deg$^{-2}$, 
or between 58 to 84 percent of the $2-10$\ keV XRB, 
depending on whether we assume the {\it HEAO1} value of 
$1.6\times 10^{-11}$\ erg\ cm$^{-2}$\ s$^{-1}$\ deg$^{-2}$
(\markcite{marshall80}Marshall et al.\ 1980) or more recent higher
estimates of $2.3\times 10^{-11}$\ erg\ cm$^{-2}$\ s$^{-1}$\ deg$^{-2}$
(e.g., \markcite{vecchi99}Vecchi et al.\ 1999). Throughout the rest 
of the paper we will conservatively adopt the 
\markcite{vecchi99}Vecchi et al.\ (1999) value. 

If we include 6 of the 7 spectroscopically unidentified sources
in the $z<2$ population (source 13 is the only source not
seen in the $B$-band and hence is the only very high redshift 
($z\gg 4$) candidate), then 56\% of the light arises from $z<2$ 
sources. If we instead restrict to the spectroscopic 
sample, then 34\% of the light arises from the $z<2$ 
population. The four sources with
spectroscopic redshifts $z>2$ contribute 13\% to the light, and
source 13 contributes just 1\%.

\section{Optical and Near-infrared Properties of the Hard 
X-ray Sample}
\label{secopt}

\subsection{Magnitudes}

At the time of this writing,
all but one (source 13) of the hard X-ray sources was detected above
the $2\sigma$ level in the NIR image, all but two (13, 16) in the $I$-band 
image, and all but one (13) in the $B$-band image. 
Source 13 has since been detected in the NIR using the CISCO
infrared camera on the Subaru 8\ m telescope 
(\markcite{cowie01}Cowie et al.\ 2001).

\begin{figure*}[tb]
\centerline{\psfig{figure=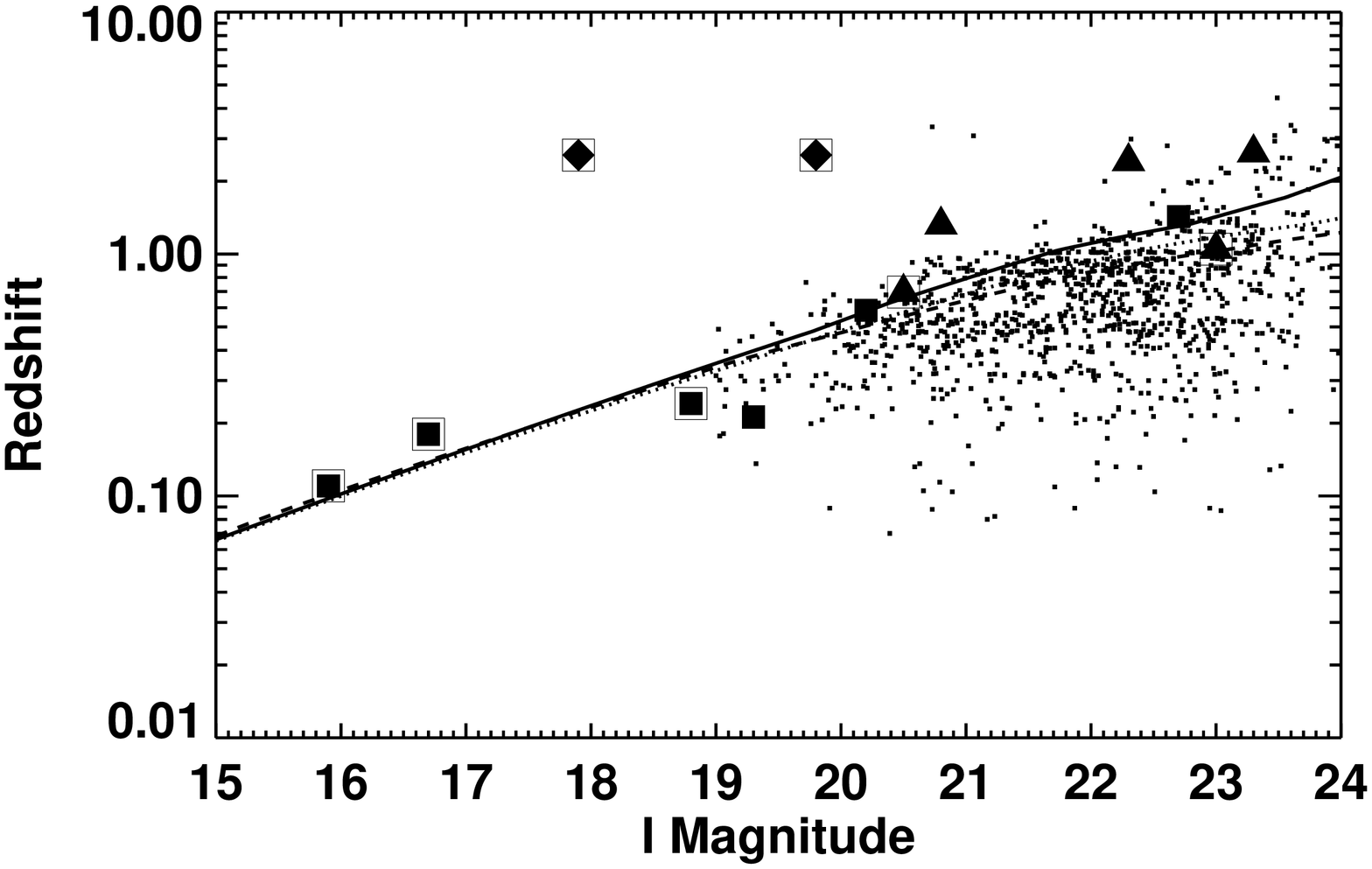,width=3.5in,angle=90}}
\figurenum{7}
\figcaption[]{
Redshift versus $I$ magnitudes for the thirteen $I<23.5$
hard X-ray sources with spectroscopic identifications (large
symbols) and for the $I<24$ field galaxies from the Hubble
Deep Field and the Hawaii Deep Survey Fields
SSA13 and SSA22 (small symbols).
Quasars are denoted by filled diamonds, AGN
by filled triangles, and optically `normal' galaxies by filled
squares. Sources with $5\sigma$
radio counterparts are enclosed in open squares.
The hard X-ray sources are almost exclusively galaxies lying at
or above $L_\ast$, as indicated by the Coleman, Wu, \& Weedman (1980)
tracks for an early-type galaxy (solid),
an early spiral galaxy (dashed), and an irregular galaxy (dotted)
with $M_{I}=-22.5$.
\label{fig7}
}
\end{figure*}

Figure~\ref{fig7} shows the redshift versus $I$ magnitude
distribution of our spectroscopically identified 
$I<23.5$ hard X-ray sample (large symbols)
and, for comparison, an optically selected $I<24$ field 
galaxy sample (small symbols). 
Hard X-ray sources with $5\sigma$ radio detections are
indicated by surrounding open boxes. The tracks are
from \markcite{coleman80}Coleman, Wu, \& Weedman (1980)
for an early-type galaxy (solid), an early spiral galaxy
(dashed), and an irregular galaxy (dotted)
with absolute magnitudes $M_I=-22.5$ in the 
assumed cosmology. The `normal' galaxies and AGN follow
the upper envelope of the star forming field galaxy population.
Thus, the hard X-ray sources predominantly lie in the most 
optically luminous galaxies. 

\subsection{Colors}
\label{seccolors}

\begin{figure*}[tb]
\centerline{\psfig{figure=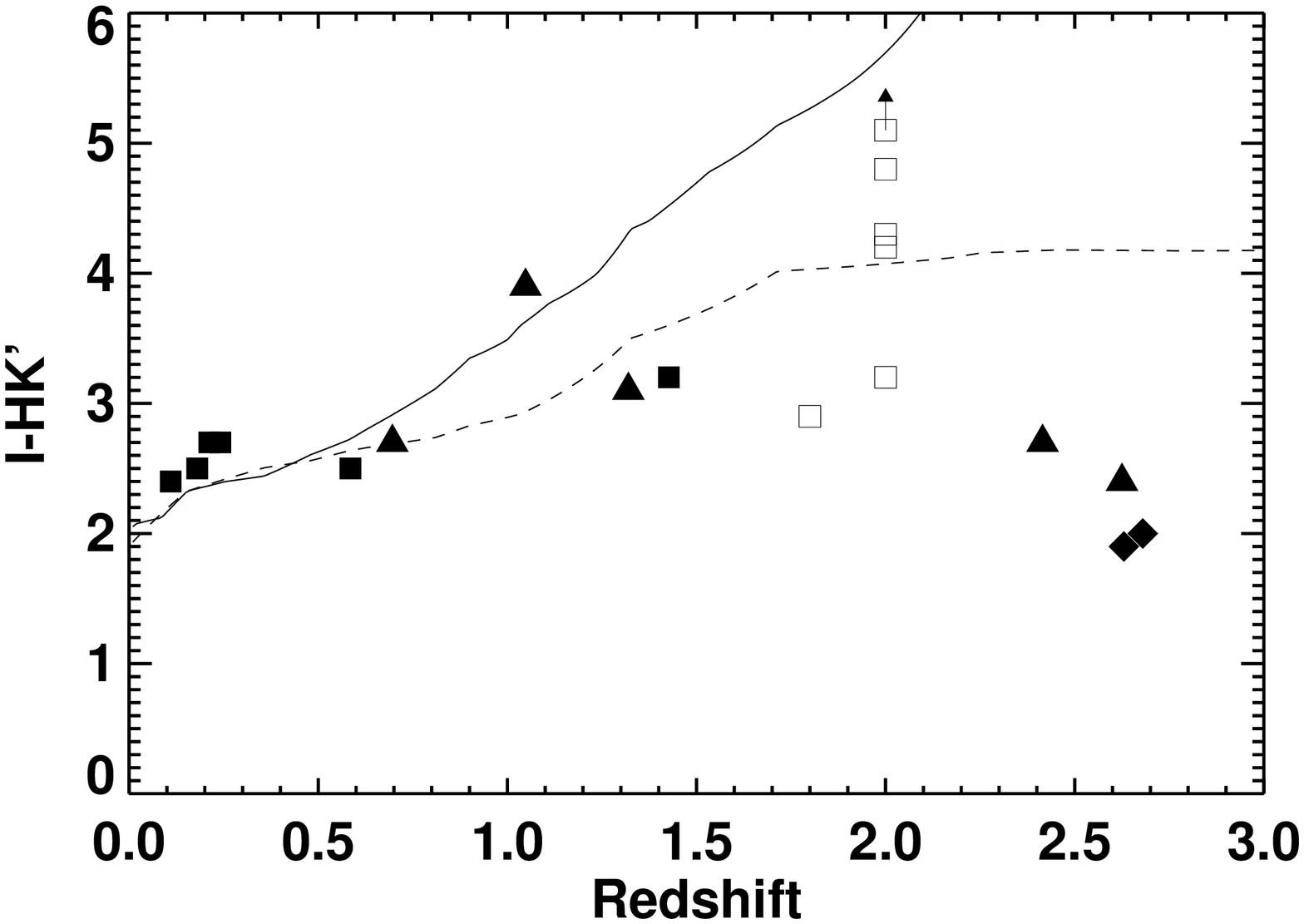,width=3.5in,angle=90}
\psfig{figure=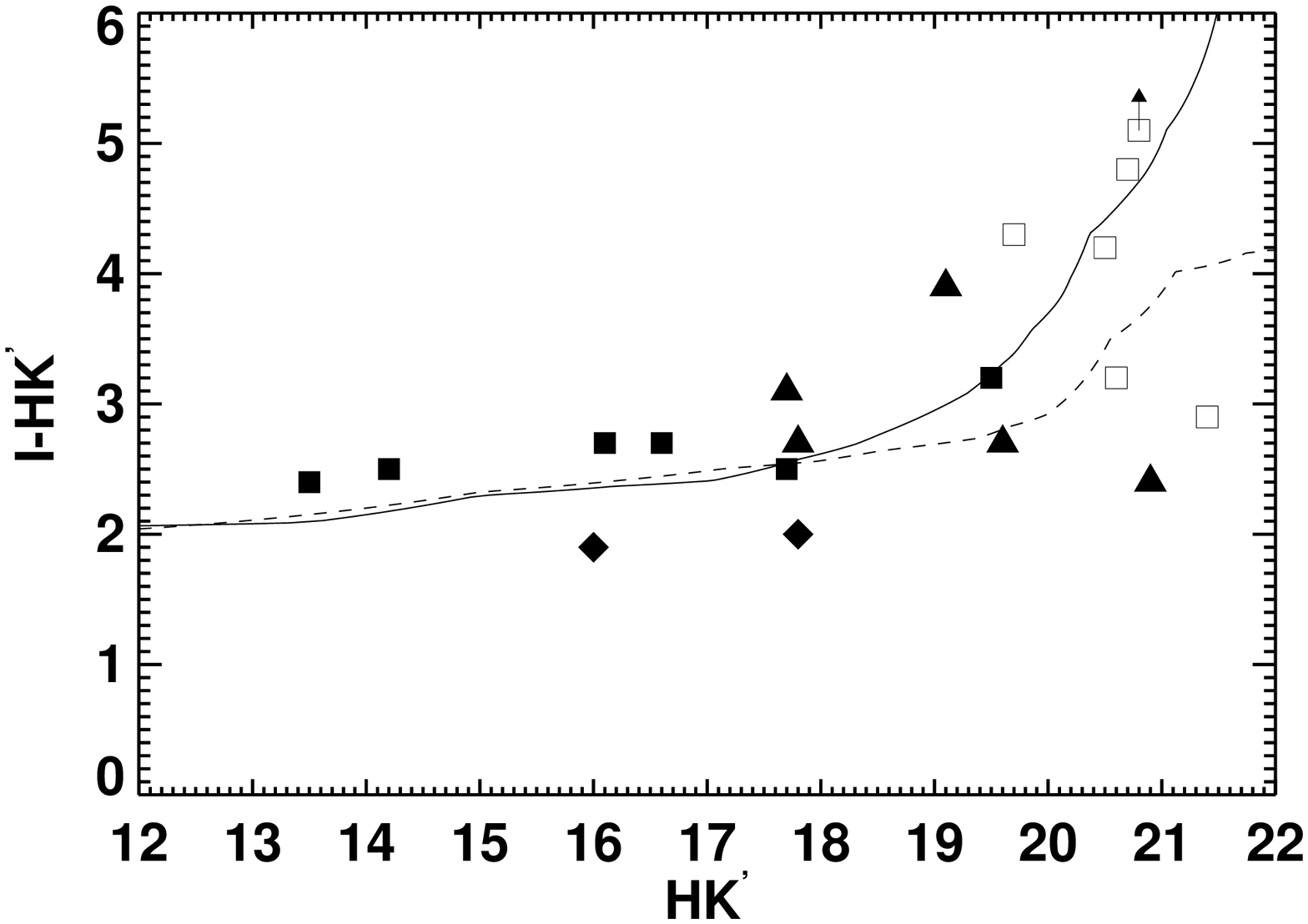,width=3.5in,angle=90}}
\figurenum{8}
\figcaption[]{
(a)\ $I-HK'$ colors versus redshift
and (b)\ $I-HK'$ colors versus $HK'$ magnitudes for the 19 hard
X-ray sources with optical and/or NIR detections
(source 13 is not shown). Quasars are denoted by filled
diamonds, AGN by filled triangles, `normal' galaxies by
filled squares, and spectroscopically unidentified sources
by open squares.
The latter, excluding source 8, are placed at $z=2$;
source 8 is placed at its millimetric redshift of 1.8.
The overlays are the
Coleman, Wu, \& Weedman (1980) tracks for an early-type
galaxy (solid curve) and an early spiral galaxy (dashed curve)
with $M_{HK'}=-25.0$. The spectroscopically unidentified sources
are likely early galaxies at $z>1.5$ because of their red colors.
\label{fig8}
}
\end{figure*}

Figures~\ref{fig8}a, b show $I-HK'$ versus redshift
and $I-HK'$ versus $HK'$ for 19 of the 20 hard X-ray sources
(source 13 is excluded).
The overlays are Coleman, Wu, \& Weedman 1980
tracks for an early-type galaxy
(solid curve) and an early spiral galaxy (dashed curve)
with $M_{HK'}=-25.0$.
The colors of the spectroscopically identified $z<2$ galaxies 
are in the range of the galaxy tracks; this is also
consistent with their morphological appearance 
(see Fig.~\ref{fig2}).
Of the sources that were too optically faint for spectroscopic 
identification, the NIR magnitudes and colors for four 
(sources 0, 9, 14, 16) suggest that they are early galaxies in 
the $z>1.5$ redshift range (Fig.~\ref{fig8}a).
\markcite{crawford00}Crawford et al.\ (2000) also
found a number of sources of this type in a sample of 
{\it Chandra} hard X-ray sources.
We were able to use radio and submillimeter detections for one 
of the optically faint sources (source 8) to estimate a millimetric 
redshift with central value $z=1.8$ (see \S~\ref{secdetection}).
We used radio detections and
submillimeter $1\sigma$ limits on sources 0, 9, 14, 16, and 17 to
estimate millimetric redshift upper limits of 
$z=$2.5, 3.0, 2.4, 2.2, and 1.4, 
respectively. Sources 8 and 17 are somewhat bluer than the curves in
Figs.~\ref{fig8}, but their AGN might be contributing substantially to 
the rest-frame optical light. 

\subsection{Bolometric Ultraviolet/Optical Luminosities}

\begin{figure*}[tb]
\centerline{\psfig{figure=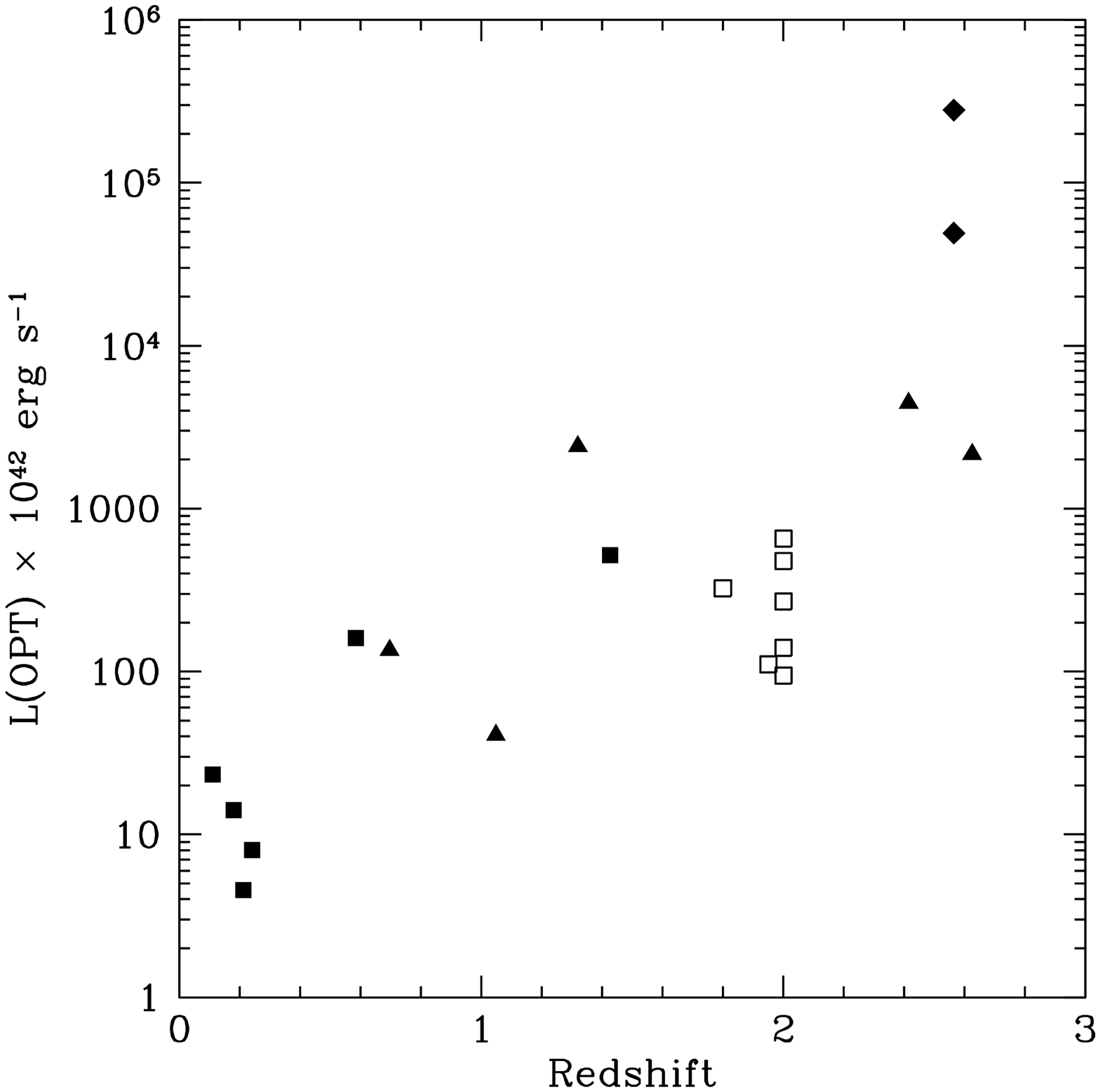,width=3in,height=3in}}
\figurenum{9}
\figcaption[]{
UV/optical luminosities versus redshift. Quasars are denoted by
filled diamonds, AGN by filled triangles, `normal' galaxies by
filled squares, and spectroscopically unidentified sources
by open squares. The latter, excluding source 8, are placed at
$z=2$; source 8 is placed at its millimetric redshift of 1.8.
Source 16 has been slightly offset
in redshift for clarity.
\label{fig9}
}
\end{figure*}

We estimate the AGN luminosities in the ultraviolet (UV)/optical 
by adopting a shape appropriate to the radio quiet AGN (e.g.,
\markcite{zheng97}Zheng et al.\ 1997). 
At frequencies below the rest-frame Lyman limit we take the spectrum
to be a $-0.8$ power-law; this steepens to a $-1.7$ power-law at higher
frequencies. Normalizing to the observed flux at the wavelength 
corresponding to the rest wavelength 2500~\AA, $f_{2500(1+z)}$, 
the UV/optical luminosity is then

$$L_{OPT} = 4\pi\ d_L^2\ (9.4\times 10^{15})\ f_{2500(1+z)}\ (1+z)^{-1}$$

\noindent
where $d_L$ is the luminosity distance in cm and 
$f_{2500(1+z)}$ is in units erg\ cm$^{-2}$\ s$^{-1}$\ Hz$^{-1}$.
We have estimated $f_{2500(1+z)}$ for the hard X-ray sample
by interpolating from the observed
fluxes corresponding to the magnitudes given in Table~1 and,
where available, the $U'$ magnitudes at 3400\ \AA\ from MCBA. 
For the more extended objects we have used the corrected $3''$ 
diameter magnitudes from MCBA rather than the $20''$ diameter 
magnitudes of Table~1 to obtain better limits on the AGN fluxes. 
Many of the
sources may still have substantial galaxy light contamination,
so $L_{OPT}$ is strictly an upper limit on the AGN
contribution. The inferred bolometric UV/optical luminosities 
are given in column~4 of Table~\ref{tab2} and are plotted
versus redshift in Fig.~\ref{fig9}.

The four sources with spectroscopic redshifts beyond $z=2$
are by far the most optically luminous sources in the sample. 
These are the two quasars (sources 3 and 6)
and the two AGN (sources 11 and 15) with narrow
Ly$\alpha$ and CIV lines. The AGN are likely dominating the
light in these sources at the observed optical and NIR wavelengths, 
as the observed colors are quite blue (see Fig.~\ref{fig8}a).

\section{An Optically Selected Sample: Hard X-ray Properties and 
Contribution to the Hard XRB}
\label{secoptsample}

The foregoing section presented the optical nature of the hard X-ray
sources. We now invert the approach and ask what are the hard X-ray
properties of an optically selected sample. In particular, we would
like to know what fraction of optical sources are significant
contributors to the hard XRB and whether these are
drawn from a particular subsample of the optical population.

\begin{figure*}[tb]
\centerline{\psfig{figure=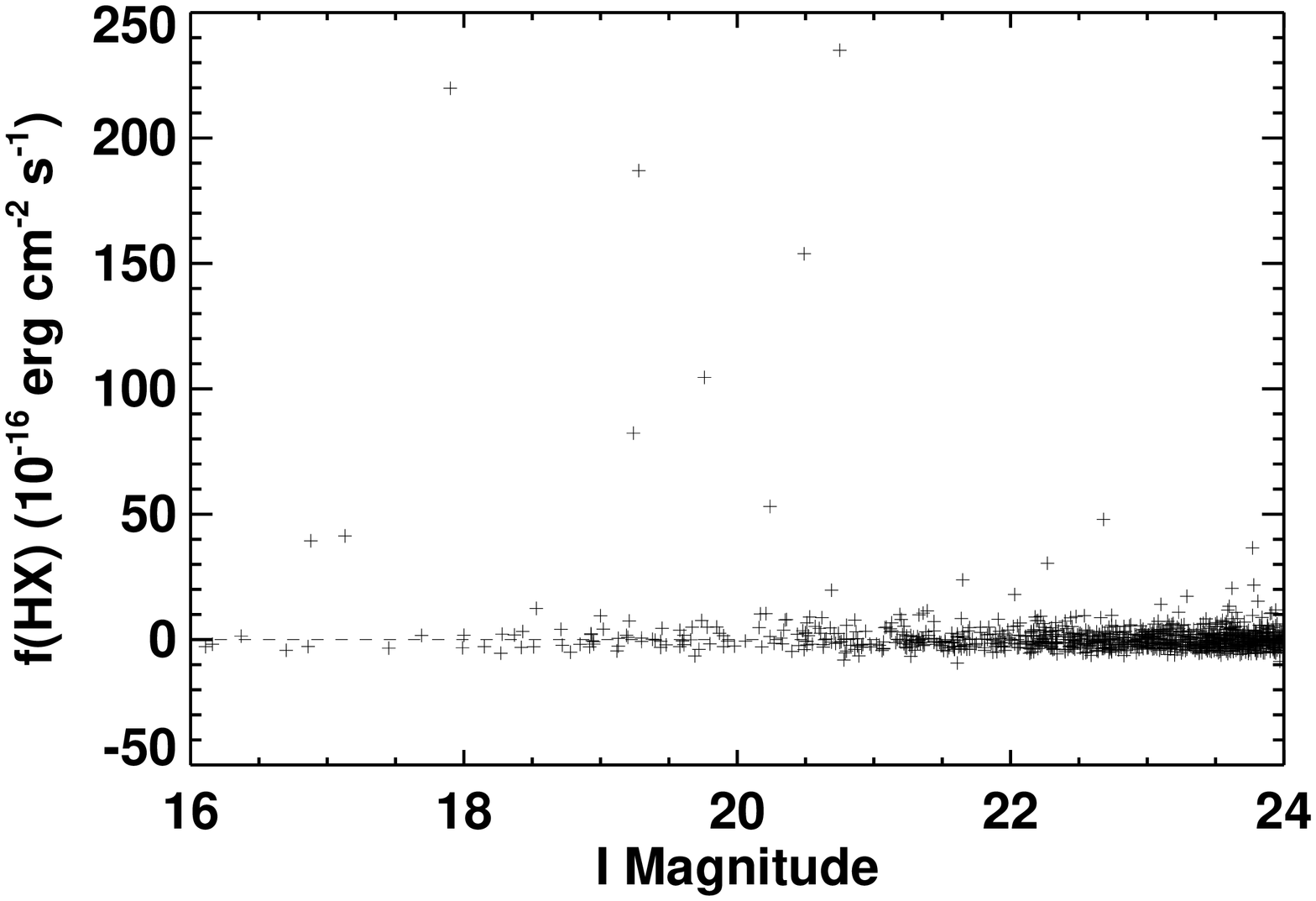,width=3.5in,angle=90}
\psfig{figure=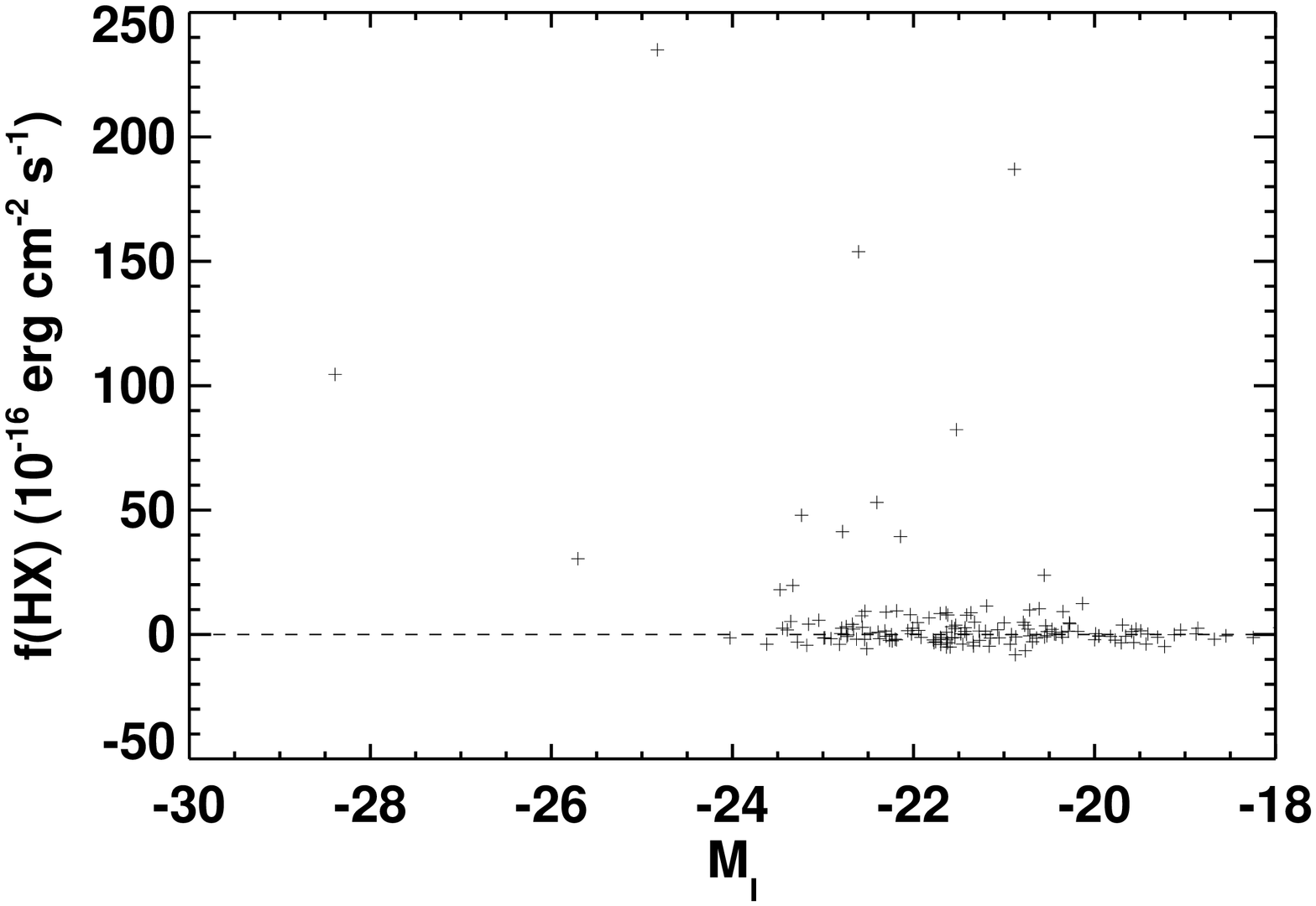,width=3.5in,angle=90}}
\figurenum{10}
\figcaption[]{
(a)\ Hard X-ray fluxes versus $I$ magnitudes for the 1151
$I<24$ galaxies and stars within a $4'$ radius of the optical
axis. Above a hard X-ray flux limit
of $2.5\times 10^{-15}$\ erg\ cm$^{-2}$\ s$^{-1}$, the
twelve $I<24$ sources of Table~1 that are within this radius
are recovered.
(b)\ Hard X-ray fluxes versus absolute $I$-band magnitudes
for the 172 $I<23$ spectroscopically identified sources
within the $4'$ radius. The K-corrections have been
computed using the Coleman, Wu, \& Weedman (1980)
SED for an early spiral galaxy.
Five of the 25 galaxies with $-22.5<M_I<-24$ have
$f_{HX}>1.0\times 10^{-15}$\ erg\ cm$^{-2}$\ s$^{-1}$.
None of the 29 galaxies with $M_I$ fainter than $-20$
is an X-ray source at this level.
\label{fig10}
}
\end{figure*}

To address this issue, we use
the complete subsample of 1151 $I<24$ galaxies and stars
within a $4'$ radius of the optical axis
of the {\it Chandra} pointing. 
The smaller radius was chosen to minimize aperture
corrections to the X-ray counts so that the X-ray
limits would be more uniform. For each of these sources, we
extracted the X-ray counts in a $2''$ radius aperture
centered on the nominal optical position and converted them to
$2-10$\ keV fluxes using the procedure outlined in \S~\ref{secdata}.
The X-ray fluxes determined in this way are plotted versus
$I$ magnitude in Fig.~\ref{fig10}a.

Above a hard X-ray flux limit of 
$2.5\times 10^{-15}$\ erg\ cm$^{-2}$\ s$^{-1}$,
we recover the twelve $I<24$ sources of Table~1 that are
within the $4'$ radius. 
Monte Carlo simulations using an equal number of sources 
at random positions show that with this extraction aperture and 
flux limit, there will be, on average, 1.4 spurious 
cross-identifications of optical sources with X-ray sources 
($<3$ at the 95 percent confidence limit). Allowing
for this contamination, we find that a fraction 
$0.009^{+0.004}_{-0.003}$
of the optical sources are X-ray sources, where the
uncertainties correspond to the 68 percent confidence ranges. 
At a lower hard X-ray flux limit of
$1.0\times 10^{-15}$\ erg\ cm$^{-2}$\ s$^{-1}$, we find 32 
sources and expect an average contamination of 10.3
($<16$ at the 95 percent confidence limit). Here, a fraction
$0.019\pm 0.005$ of the optical sources are X-ray sources.

Excluding the incompletely covered edges of the S3 and S2
chips, the observed area is 47\ arcmin$^2$, 
and the contribution to the $2-10$\ keV background of the 12 sources 
above the $2.5\times 10^{-15}$\ erg\ cm$^{-2}$\ s$^{-1}$ 
hard X-ray flux limit is 
$9.5\times 10^{-12}$\ erg\ cm$^{-2}$\ s$^{-1}$\ deg$^{-2}$
(41\% of the hard XRB).

If we assume that half of the sources in the flux
range $1.0\times 10^{-15}$\ erg\ cm$^{-2}$\ s$^{-1}$ to
$2.5\times 10^{-15}$\ erg\ cm$^{-2}$\ s$^{-1}$
are real, we estimate a further contribution of only
$1.1\times 10^{-12}$\ erg\ cm$^{-2}$\ s$^{-1}$\ deg$^{-2}$;
thus, increasing the X-ray sensitivity adds little to the
hard XRB contribution. 
Indeed, summing over all the optically selected
sources above $1.0\times 10^{-15}$\ erg\ cm$^{-2}$\ s$^{-1}$
yields a total of
$10.4 \pm 0.8 \times 10^{-12}$\ erg\ cm$^{-2}$\ s$^{-1}$\ deg$^{-2}$,
where the uncertainties are the 68 percent confidence ranges
based on randomized samples. 

We conclude that most of the hard XRB from the optically 
selected sources is dominated by 
the small number of sources with hard X-ray fluxes above
$2.5\times 10^{-15}$\ erg\ cm$^{-2}$\ s$^{-1}$,
as is also the case for the directly selected hard
X-ray sample.

Another way to estimate the fraction of optically selected sources
that are X-ray sources is to use a spectroscopic data sample.
Of the 554 $I<23$ objects in the $4'$ radius region, 172 have
measured redshifts or are spectroscopically identified stars.
The measured $2-10$\ keV fluxes of these galaxies are shown versus
their absolute $I$-band magnitudes, $M_I$, in Fig.~\ref{fig10}b. 
Here K-corrections have been computed using the
\markcite{coleman80}Coleman, Wu \& Weedman (1980) spectral energy
distribution (SED) for an early spiral galaxy.

Of the 25 galaxies with $-22.5<M_I<-24$, five have hard X-ray fluxes 
above $1.0\times 10^{-15}$\ erg\ cm$^{-2}$\ s$^{-1}$.
By contrast, none of the 29 galaxies with $M_I$ fainter
than $-20$ is an X-ray source at this level.
If we correct for the incompleteness of the spectroscopically 
identified optical sample, we find that the 
fraction of optically luminous galaxies that are X-ray sources is
$0.07^{+0.05}_{-0.03}$. Thus, a very substantial fraction of the 
optically luminous galaxies are undergoing X-ray activity at any 
given time.

\section{Submillimeter and Radio Properties of the Hard X-ray Sample}
\label{secsmmradio}

The new population of highly obscured, exceptionally luminous
sources discovered by SCUBA 
(\markcite{smail97}Smail, Ivison, \& Blain 1997;
\markcite{hughes98}Hughes et al.\ 1998;
\markcite{barger98}Barger et al.\ 1998; 
\markcite{bcs99}Barger, Cowie, \& Sanders 1999; 
\markcite{eales99,eales00}Eales et al.\ 1999, 2000)
appear to be distant analogs of the local ultraluminous infrared 
galaxies (ULIGs; \markcite{sanders96}Sanders \& Mirabel 1996).
There is an ongoing debate on whether local ULIGs are dominantly 
powered by star formation or by heavily dust enshrouded AGN,
and the same applies to the distant SCUBA sources
(see, e.g., \markcite{trentham00}Trentham 2000). 
If the hard X-ray sources are highly absorbed,
then the rest-frame soft X-ray through NIR radiation will be
reprocessed by dust and gas, and the energy will appear in
the FIR. At high redshifts ($z\gg 1$) the FIR radiation is
shifted to the submillimeter.

\markcite{barger99}Barger et al.\ (1999) carried out a 
spectroscopic survey
of a complete sample of submillimeter sources detected in
a survey of massive lensing clusters. Only three of the 17 sources
could be reliably identified spectroscopically; of these,
two showed AGN signatures. Thus, the possibility that most
SCUBA sources contain AGN remains open. Several authors 
(\markcite{almaini99}Almaini, Lawrence, \& Boyle\ 1999;
\markcite{gunn00}Gunn \& Shanks 2000)
have modelled the X-ray and submillimeter backgrounds; they
predict an AGN contribution to the SCUBA surveys
at the level of $10-20$ percent.

The results of recent searches for submillimeter counterparts 
to {\it Chandra} X-ray sources have been mixed. In a study of
two massive lensing clusters, A2390 and A1835,
\markcite{fabian00}Fabian et al.\ (2000) 
identified three significant $2-7$\ keV sources,
but these were not significantly detected in the submillimeter
(\markcite{smail98}Smail et al.\ 1998).
Likewise, \markcite{horn00}Hornschemeier et al.\ (2000)
did not see either of their $2-8$\ keV sources in the ultradeep
Hubble Deep Field SCUBA map of
\markcite{hughes98}Hughes et al.\ (1998).
In contrast, both of the $2-10$\ keV sources
detected by \markcite{bautz00}Bautz et al.\ (2000) in the A370 
lensed field are submillimeter sources 
(\markcite{smail97}Smail, Ivison, \& Blain\ 1997).
These two sources were previously identified spectroscopically
as AGN (\markcite{ivison98}Ivison et al.\ 1998;
\markcite{barger99}Barger et al.\ 1999). 
The above mixed results probably reflect the fact that the 
850\ $\mu$m flux limits obtainable with SCUBA are quite close 
to the expected fluxes from the obscured AGN, as we address below.

The present study has the advantage that wide-area submillimeter 
(\markcite{bcs99}Barger, Cowie, \& Sanders 1999 and the present paper)
and extremely deep 20\ cm data 
(\markcite{richards01}Richards et al.\ 2001)
exist over the entire X-ray field of 57\ arcmin$^2$. 
An ultradeep 50 hr submillimeter map
(\markcite{barger98}Barger et al.\ 1998) also exists for one region 
of the field that contains two hard X-ray sources.

\subsection{Submillimeter Detection of a Hard X-ray Source}
\label{secdetection}

With one exception, the hard X-ray sources are not detected in the
submillimeter at the $3\sigma$ level.
The exception is source 8, an optically faint,
highly absorbed ($\Gamma=-0.06$) X-ray source
in the ultradeep submillimeter map.

The bolometric FIR flux is related to the rest-frame 20\ cm flux 
through the well-established FIR-radio correlation
(\markcite{condon92}Condon 1992) of local starburst galaxies and
radio quiet AGN. The SEDs of submillimeter sources are
reasonably well approximated by the thermal black-body spectrum
of the ULIG Arp~220
(\markcite{cy00}Carilli \& Yun 2000;
\markcite{bcr00}Barger, Cowie, \& Richards 2000), which is powered
by star formation (\markcite{downes98}Downes \& Solomon\ 1998).
We can therefore use the submillimeter to
radio flux ratio to infer an approximate redshift of 1.8
(with estimated redshift range $1.2-2.4$) for source 8 (see
Eqs.~2 and 4 of \markcite{bcr00}Barger, Cowie, \& Richards 2000).
The millimetric redshift estimation technique is expected
to hold for sources dominated by star formation
or for radio quiet AGN.

\subsection{Spectral Energy Distributions of the Hard X-ray Sources}
\label{secsed}

\begin{figure*}[tb]
\centerline{\psfig{figure=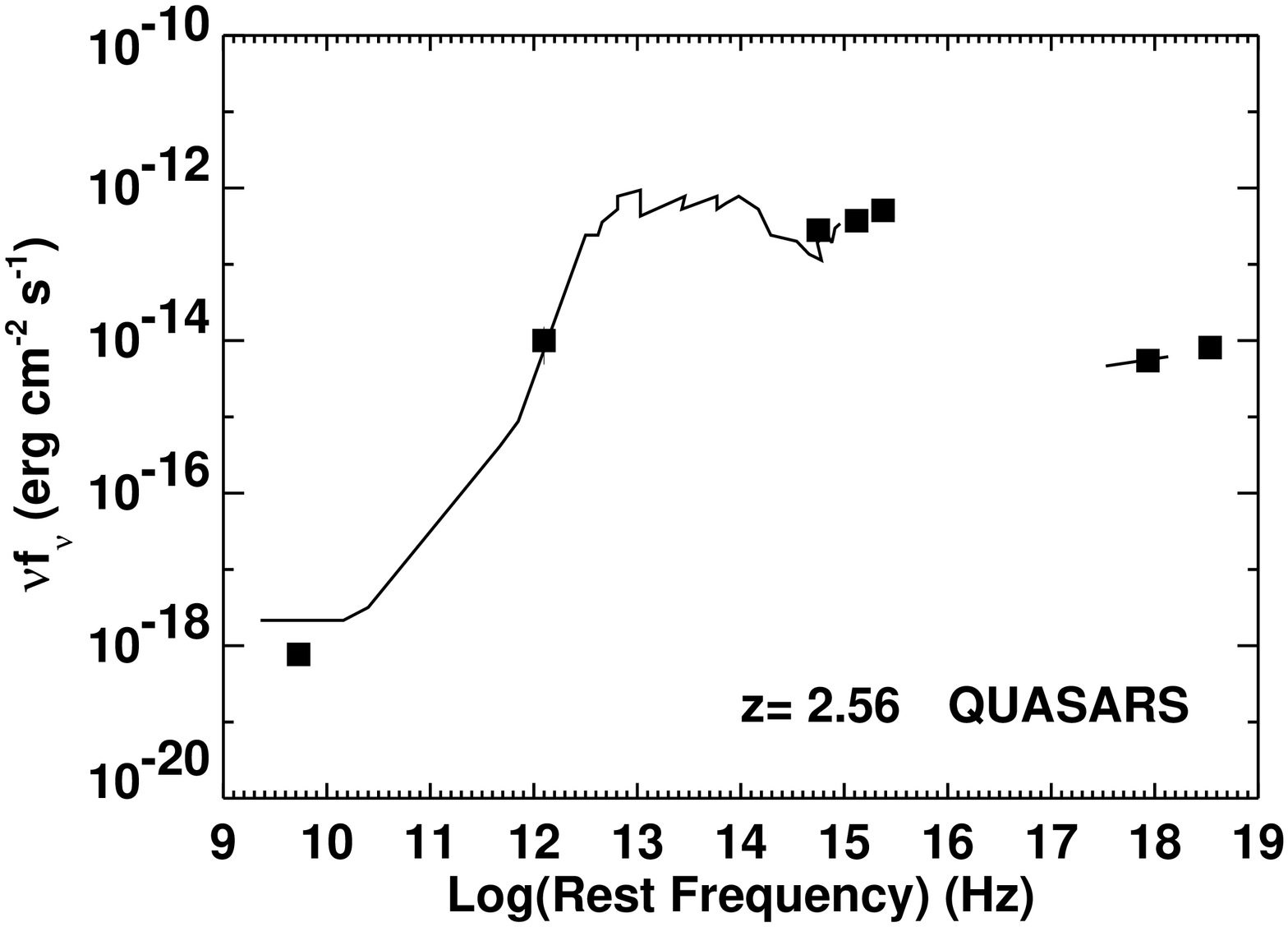,width=3.5in,angle=90}
\psfig{figure=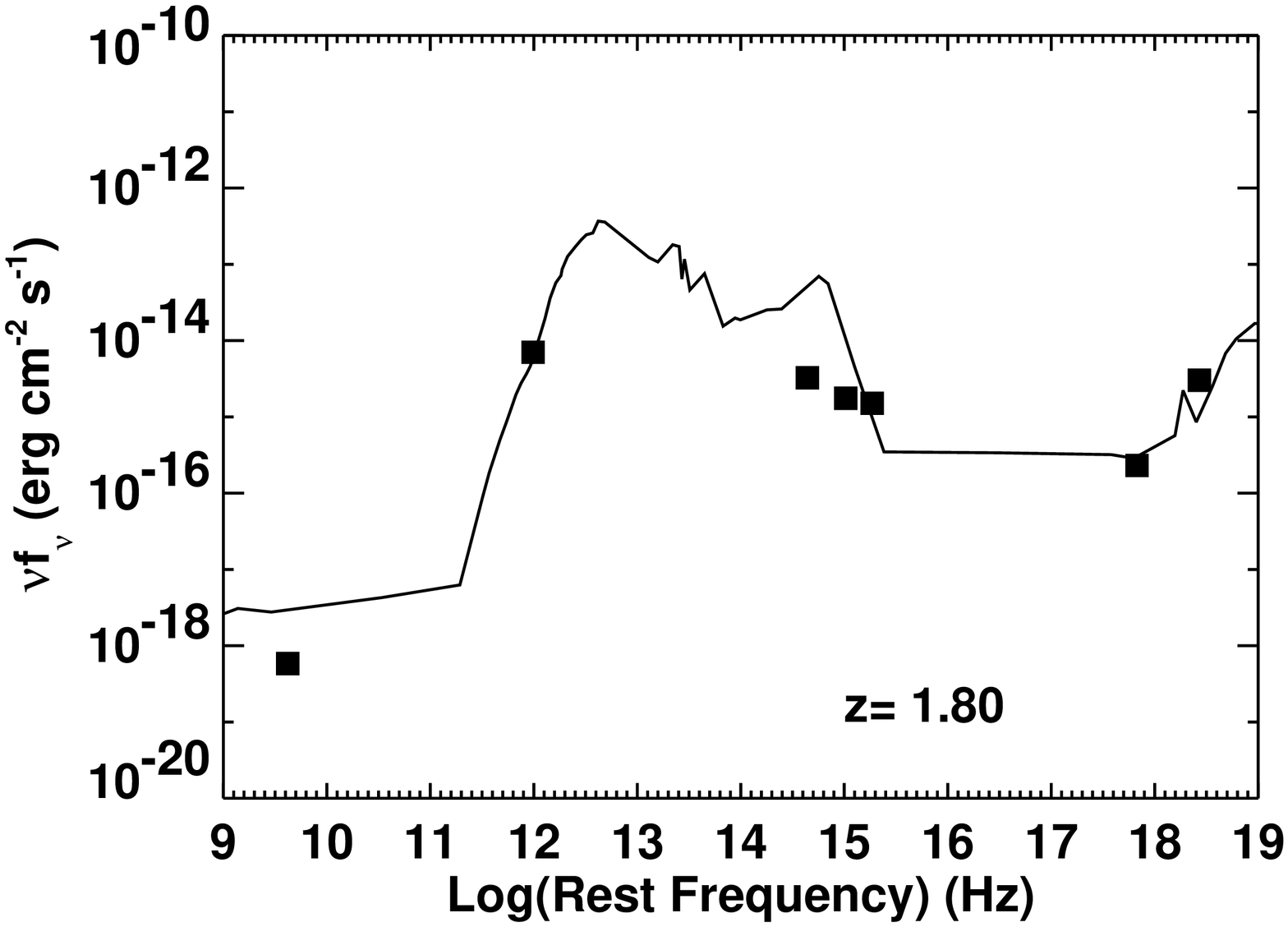,width=3.5in,angle=90}}
\figurenum{11}
\figcaption[]{
(a)\ The average rest-frame SED ($\nu f_\nu$) of the two $z=2.56$
quasars (filled squares) is compared with the rest-frame SED
of the radio quiet quasar PG~1543+489 (solid curve).
The radio to optical/NIR data on PG~1543+489 were taken from
the compilation in Polletta et al.\ (2000), the soft X-ray data
from Laor et al.\ (1997), and the hard X-ray data from
Vaughan et al.\ (1999).
(b)\ The rest-frame SED of source 8 (filled squares) is shown in
comparison with the SED (from Hasinger 2000) of the heavily dust
obscured local AGN NGC~6240 (solid curve).
\label{fig11}
}
\end{figure*}

The two quasars are the softest sources in the hard X-ray 
sample. In Fig.~\ref{fig11}a\ we compare their average 
rest-frame SED (filled squares) to the rest-frame SED of the 
$z=0.4$ radio quiet quasar PG~1543+489 (solid curve),
normalized to approximately match the
soft and hard X-ray data of the averaged quasars.

Most of the sources in the sample show much harder X-ray spectra 
than the two quasars and are consistent with being obscured AGN.
In Fig.~\ref{fig11}b we compare the rest-frame SED of source 8
(filled squares), based on the millimetric redshift of 1.8, with
the SED of the local heavily dust obscured AGN NGC~6240 (solid curve), 
normalized to approximately match the hard and soft X-ray 
data of source 8.

\subsection{Bolometric FIR Luminosities Inferred from Radio Fluxes}

The sources with spectroscopic identifications are mostly
at modest redshifts where radio observations
provide a more sensitive route for obtaining total FIR luminosities
than do submillimeter observations 
(\markcite{bcr00}Barger, Cowie, \& Richards 2000).
Assuming that the sources are well described by the FIR-radio 
correlation (which holds for both star forming and radio-quiet AGN)
given in \markcite{sanders96}Sanders \& Mirabel (1996) with $q=2.35$, 
then $f_{FIR}=8.4\times 10^{14}\ f_{20}$, where 
$f_{20}$ is the flux at 20\ cm. We calculate the FIR
luminosity by relating it to the observed 20\ cm flux,
assuming a synchrotron spectrum with an index of $-0.8$; then 

$$L_{FIR}=4\ \pi\ d_L^2\ (8.4\times 10^{14})\ f_{20}\ (1+z)^{-0.2}$$

\noindent
where $d_L$ is the luminosity distance in cm and
$f_{20}$ has units erg\ cm$^{-2}$\ s$^{-1}$\ Hz$^{-1}$.

The bolometric FIR luminosities obtained by this method are given 
in column~5 of Table~\ref{tab2} and are plotted in Fig.~\ref{fig12} 
as open circles. Four sources are not detected above the 
15\ $\mu$Jy ($3\sigma$) limit in the 20\ cm data and their 
luminosities are given as upper limits based on the $3\sigma$ 
radio limit. For our sources without spectroscopic or millimetric
redshifts, we nominally assume $z=2.0$.
It is interesting to note that three-quarters of the hard X-ray 
sources with radio detections in our sample would be classified as LIGs
($L_{FIR}>10^{11}\ L_\odot$; 
\markcite{sanders96}Sanders \& Mirabel 1996)
from their FIR luminosities, and half would be classified as ULIGs
($L_{FIR}>10^{12}\ L_\odot$).

\subsection{Bolometric FIR Luminosities Inferred from 
Submillimeter Fluxes}

By assuming an ULIG SED, we can use the
850\ $\mu$m data to estimate the FIR luminosity of 
source 8 and to place upper limits on the FIR luminosities of the 
remaining sources. For our non-quasar X-ray sources with redshifts, 
we estimate the FIR luminosities by scaling the NGC~6240 luminosity
by the ratio of the 850\ $\mu$m fluxes
(after placing NGC~6240 at the redshift of the source). 
For NGC~6240 we use the $T=42$\ K cold component determined by 
\markcite{klaas97}Klaas et al.\ (1997),
which has $L_{FIR}=2.0\times 10^{45}\ h_{65}$\ erg\ s$^{-1}$.
For our X-ray sources with no submillimeter
detections, we use the $3\sigma$ upper limits on the submillimeter 
fluxes in calculating the $L_{FIR}$ values. 
For our sources without spectroscopic or millimetric redshifts, 
we nominally assume $z=2.0$. Since
above $z\sim 1$ the luminosity distance dependence and the 
K-correction approximately cancel, there is a simple translation
of flux to bolometric luminosity independent of redshift; thus, 
our procedure should give accurate upper limits for the
unidentified sources as long as they are not at low redshift 
($z<1$).

For our two quasar X-ray sources, we estimate the FIR luminosities
as above using the SED and FIR luminosity of PG~1543+489.
For PG~1543+489
we calculate $L_{FIR}=10.5\times 10^{45}\ h_{65}$\ erg\ s$^{-1}$
over the range $40-500\ \mu$m using the $T=46$\ K grey-body model
given in \markcite{polletta}Polletta et al.\ (2000).

\begin{figure*}[tb]
\centerline{\psfig{figure=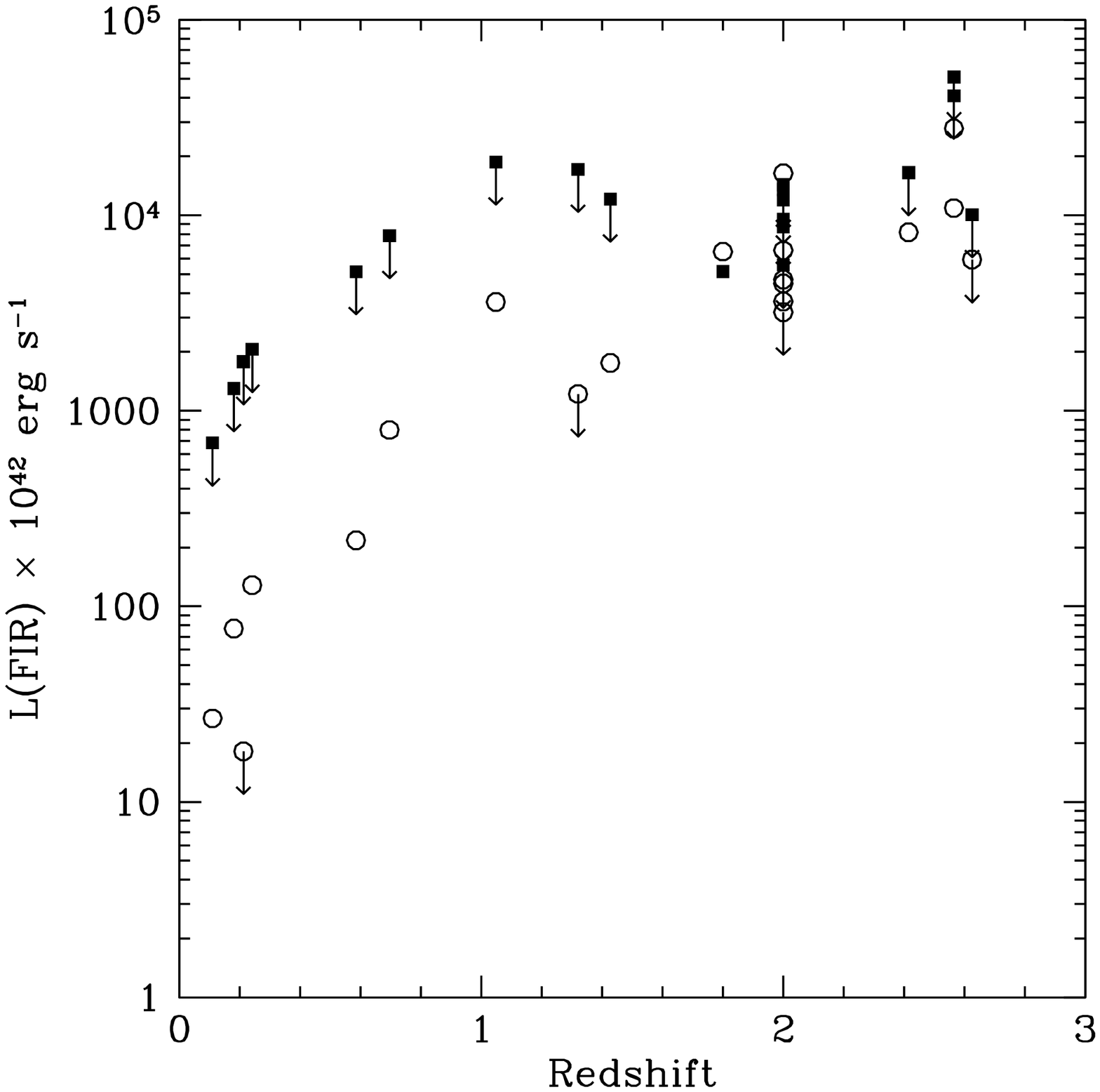,width=3in,height=3in}}
\figurenum{12}
\figcaption[]{
FIR luminosities versus redshift for the hard X-ray sample.
Open circles were calculated from the radio data and the
FIR-radio correlation. Filled squares were calculated from the
submillimeter data and the NGC~6240 or PG~1543+489 SEDs
and luminosities (see text for details).
The spectroscopically unidentified sources, excluding
source 8, are placed at $z=2$; source 8 is placed at its
millimetric redshift of 1.8.
Downward pointing arrows indicate $3\sigma$ limits for either
the radio inferred FIR luminosities or
the submillimeter inferred FIR luminosities.
\label{fig12}
}
\end{figure*}

The bolometric FIR luminosities or upper limits 
obtained from the submillimeter fluxes in this way are given in 
column~6 of Table~\ref{tab2} and are plotted in Fig.~\ref{fig12} as filled 
squares. The radio inferred FIR luminosities or upper limits 
are consistent with the submillimeter inferred FIR luminosities 
or upper limits. It is clear from Fig.~\ref{fig12} that
more sensitive submillimeter observations would be
required to detect even the higher redshifts objects.

\subsection{Bolometric Luminosity Ratios}
\label{secbololumratio}

\begin{figure*}[tb]
\centerline{\psfig{figure=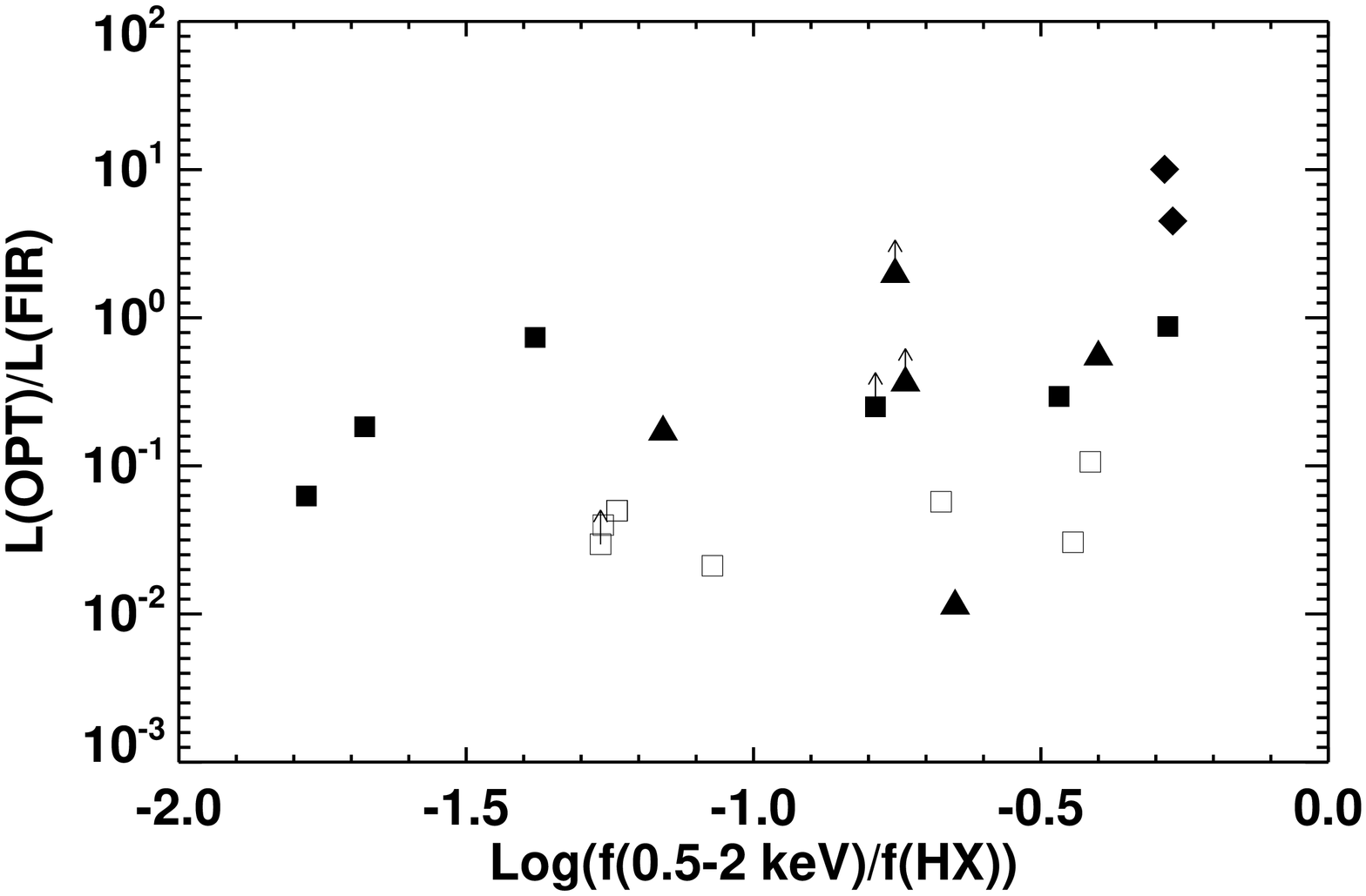,width=3.5in,angle=90}}
\figurenum{13}
\figcaption[]{
UV/optical to FIR luminosity ratios versus soft to hard
X-ray flux ratios. Quasars are denoted by filled
diamonds, AGN by filled triangles, `normal' galaxies by
filled squares, and spectroscopically unidentified sources
by open squares. The latter, excluding source 8, are placed at $z=2$;
source 8 is placed at its millimetric redshift of 1.8.
Upward pointing arrows indicate $3\sigma$ upper limits on the
radio inferred FIR luminosities.
\label{fig13}
}
\end{figure*}

In Fig.~\ref{fig13} we plot the UV/optical to FIR luminosity
ratios versus the logarithm of the soft to hard X-ray flux ratios. 
If the X-ray sources are hardened by line-of-sight opacities and 
the same material extinguishes
the UV/optical light, then as the opacities decrease, 
we would expect both the UV/optical to FIR luminosity ratios
and the soft to hard X-ray flux ratios to increase.
For the spectroscopically identified sources there
is a broad overall trend in this sense but with large scatter.

\begin{figure*}[tb]
\centerline{\psfig{figure=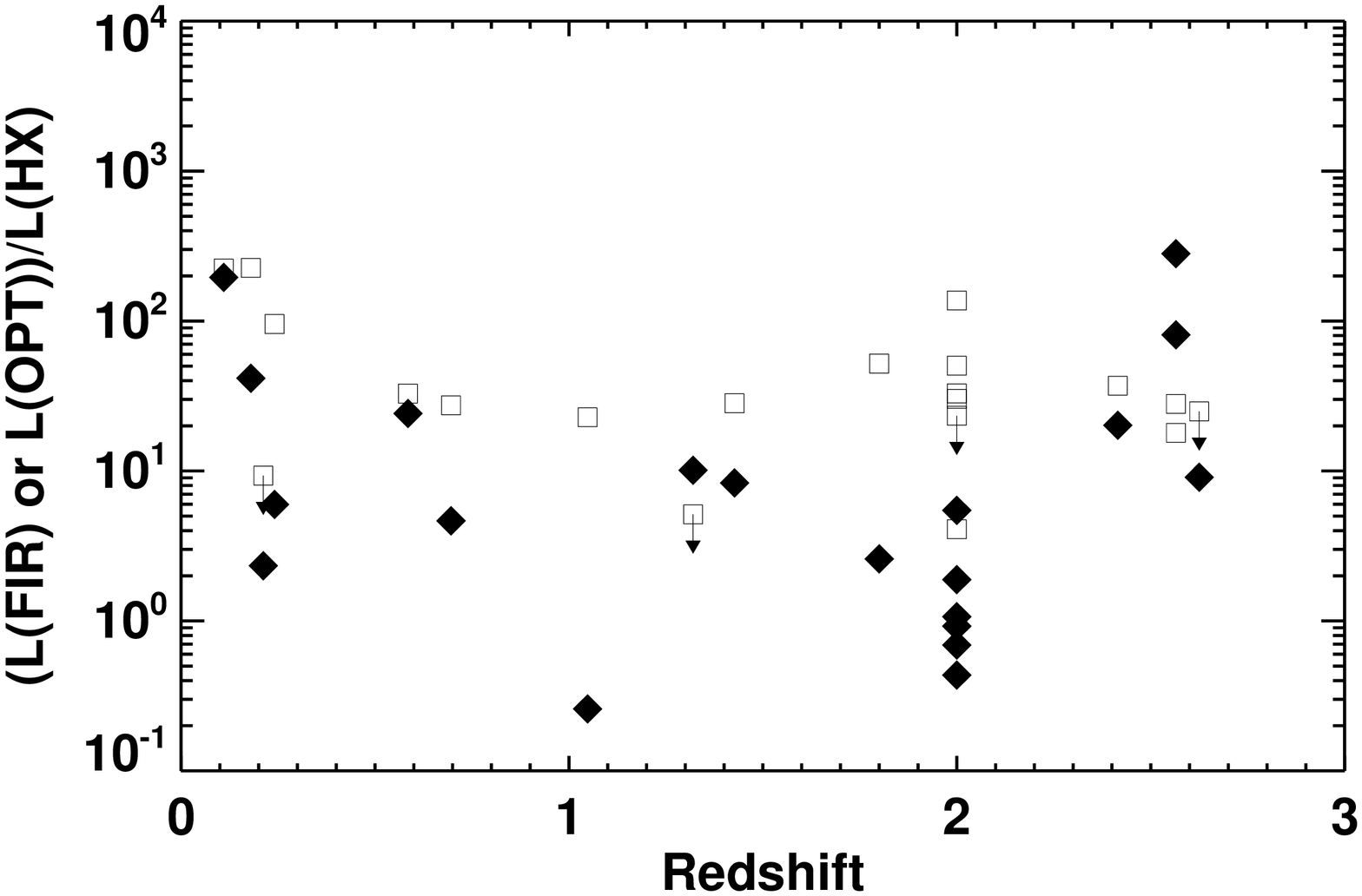,width=3.5in,angle=90}}
\figurenum{14}
\figcaption[]{
Ratios of the bolometric optical (filled diamonds) and
radio inferred FIR (open squares)
luminosities to the hard X-ray luminosities versus redshift.
Spectroscopically unidentified sources, excluding source 8,
are placed at $z=2$; source 8 is placed at its millimetric
redshift of 1.8. Downward pointing arrows indicate
$3\sigma$ limits for the radio inferred FIR luminosities.
\label{fig14}
}
\end{figure*}

In Fig.~\ref{fig14} we plot the ratios
$L_{FIR}/L_{HX}$ (open squares) and $L_{OPT}/L_{HX}$
(filled diamonds) versus redshift. 
Only the two quasars and source 2 at $z=1.320$, with its curious
optical absorption line spectrum, are dominated by the
UV/optical light rather than by the FIR light.
The two quasars are the softest sources in the sample and 
have X-ray photon indices (1.75 and 1.80) consistent
with no absorption. Source 2
has a harder photon index (0.9) than the quasars;
thus, the dominance of the UV/optical light here
is perhaps less expected and may be due to substantial 
contributions of stellar light.
For the remaining sources the FIR light dominates.

\subsection{Bolometric Correction}
\label{secbolo}

If we assume that the FIR light in the hard X-ray sources
is reprocessed AGN light, uncontaminated by star formation
in the galaxies, then we may use the present data to compute
the bolometric correction from the hard X-ray luminosity to the
bolometric light of the AGN.
Excluding the two quasars, which have an average
$(L_{FIR}+L_{OPT})/L_{HX}=200$, the remaining
sources have an average $L_{OPT}/L_{HX}=19$. 
If the sources with $3\sigma$ upper limits on $L_{FIR}$ are included 
at this level, then the average
$L_{FIR}/L_{HX}=59$, where we have used the radio inferred
FIR luminosities, and the average
$(L_{FIR}+L_{OPT})/L_{HX}=78$.
If the sources with upper limits are instead
assigned zero luminosity ratios, 
then the average $L_{FIR}/L_{HX}=56$ and the average
$(L_{FIR}+L_{OPT})/L_{HX}=74$.

The ratios are weighted to higher values by the two lowest
redshift sources. If we instead weight the averages by the hard 
X-ray fluxes, then the $L_{FIR}/L_{HX}$ ratio drops to 33,
the $(L_{FIR}+L_{OPT})/L_{HX}$ ratio to 42, 
and the $L_{OPT}/L_{HX}$ ratio to 9.
Because galaxy contamination of
the optical light is a problem, the actual bolometric correction
probably lies somewhere between the values of $L_{FIR}/L_{HX}$ and
$(L_{FIR}+L_{OPT})/L_{HX}$.
As a conservative minimum, we adopt the $L_{FIR}/L_{HX}=33$ ratio 
for the bolometric correction in subsequent discussions while 
recognizing that this ratio could be larger by a factor of two
or more.
%the factor of two or more uncertainty in this value.

The average $L_{OPT}/L_{HX}$ ratio for the hard X-ray sources is much
smaller than the $B$-band to $0.5-4$\ keV luminosity ratios
seen in local early-type galaxies, whether we consider the
soft components thought to arise from thermal emission in the
gaseous halos (luminosity ratios in the range $150-40000$)
or the hard components which may be produced
by X-ray binaries (luminosity ratios in the range $1200-8000$)
(\markcite{matsumoto97}Matsumoto et al.\ 1997).
The much higher X-ray luminosities of the present sources
relative to their stellar luminosities provide strong evidence
that they are powered by AGN.

It is also interesting to consider the total X-ray light. 
For an assumed photon index of $\Gamma=2$, 
the total X-ray light is only weakly sensitive to the adopted energy 
range. For an energy range from 0.1 to 100\ keV, the ratio of the 
total X-ray luminosity to the $2-10$\ keV luminosity is only a factor
$\ln(1000)/\ln(5)=4.3$.
The total X-ray luminosities of the sources, $L_{X}$,
are still substantially smaller than the bolometric luminosities at 
other wavelengths, but an appreciable fraction of the AGN light
(typically $1-20$\%) is emerging in the X-rays. 
$L_{BOL}=L_{FIR}+L_{OPT}+L_{X}$ values are given in the last 
column of Table~\ref{tab2}; here again the radio determinations are used as 
the $L_{FIR}$ inputs.

\subsection{FIR and Hard X-ray Luminosity Comparisons}

\begin{figure*}[tb]
\centerline{\psfig{figure=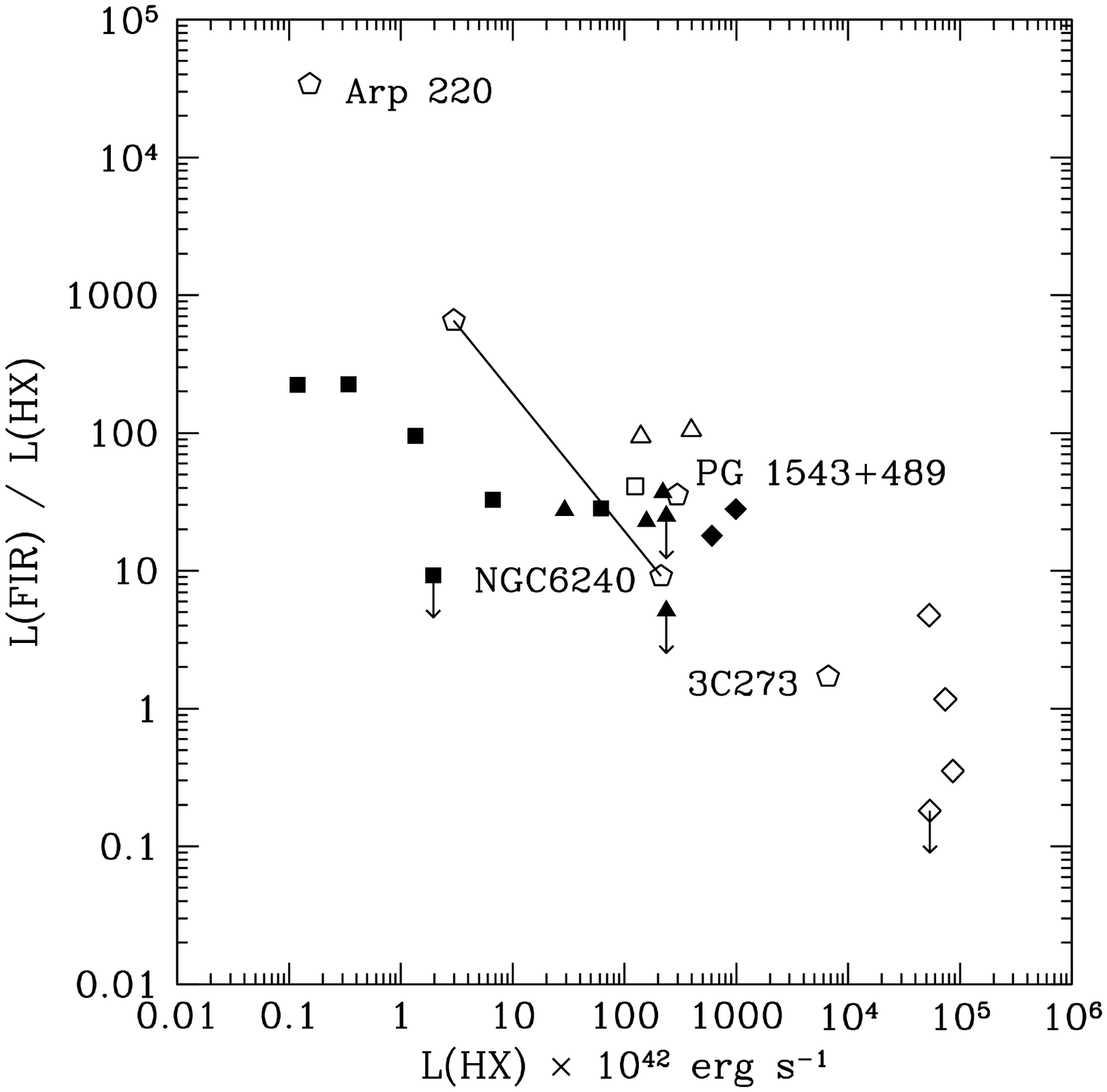,width=3in,height=3in}
\psfig{figure=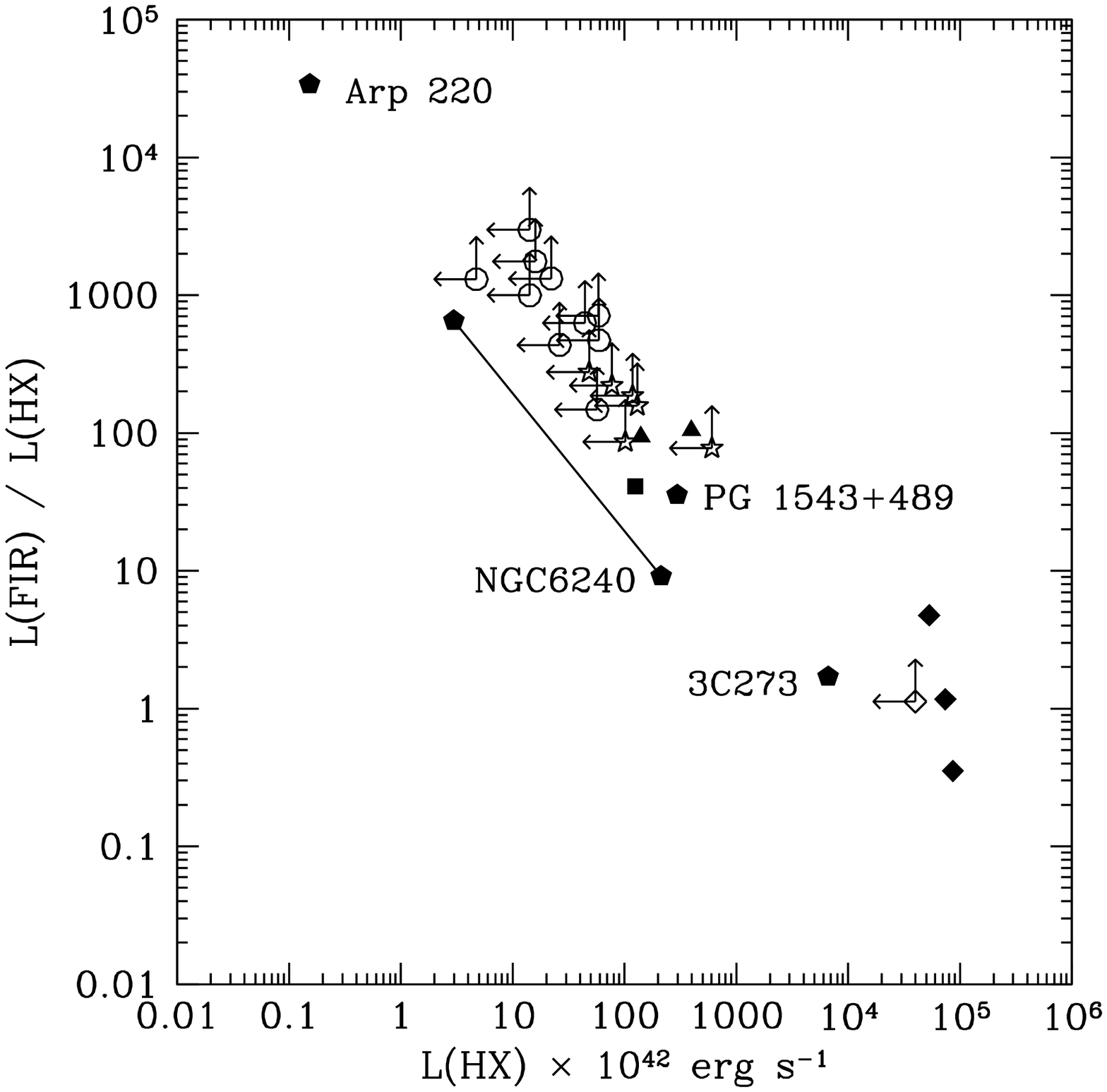,width=3in,height=3in}}
\figurenum{15}
\figcaption[]{
(a)\ $L_{FIR}/L_{HX}$ versus $L_{HX}$ for the hard X-ray sample
with redshifts.
Quasars are denoted by filled diamonds, AGN by filled triangles,
`normal' galaxies by filled squares, and the optically faint
source 8 by an open square.
Also shown in the figure (see text for references)
are the data values for the local ULIGs Arp~220
and NGC~6240, the radio quiet quasar PG~1543+489,
and the brightest quasar in the sky, 3C~273.
A straight line has been drawn between the absorption-corrected
(higher X-ray luminosity) and observed NGC~6240 ratios to
illustrate the large effects absorption can make at low redshift.
The luminous $z>4$ radio quiet quasars detected in
both the submillimeter and X-ray are shown by the open diamonds.
The two significant hard X-ray and submillimeter sources in A370 
are shown by the open triangles.
(b)\ $L_{FIR}/L_{HX}$ versus $L_{HX}$ for $>3\sigma$ submillimeter
sources with spectroscopic or millimetric redshifts
and hard X-ray detections or limits. Open circles denote
the sources in the Hubble Deep Field and Flanking Fields, and
open stars denote the sources in the SSA13 field (see text for
references). Symbols for the
comparison sources and for the sources significantly detected in
hard X-rays have been filled in for clarity.
\label{fig15}
}
\end{figure*}

By considering the ratio of the FIR luminosity, $L_{FIR}$, to the hard
X-ray luminosity, $L_{HX}$, we can eliminate any redshift or cosmology
dependence and make relative comparisons of distant and local systems.
Figure~\ref{fig15}a shows the $L_{FIR}/L_{HX}$ ratio versus 
$L_{HX}$ for our data sample.
Following \markcite{fabian00}Fabian et al.\ (2000),
we also show on the figure the data values for 
the local ultraluminous infrared galaxies 
Arp~220 and NGC~6240 (\markcite{klaas97}Klaas et al.\ 1997;
\markcite{iwasawa99}Iwasawa 1999), 
the $z=0.4$ radio quiet quasar PG~1543+489
(\markcite{polletta}Polletta et al.\ 2000;
\markcite{vaughan}Vaughan et al.\ 1999), 
and the radio loud quasar 3C~273 
(\markcite{kim98}Kim \& Sanders 1998;
\markcite{turler99}T\"{u}rler et al.\ 1999).
NGC~6240 hosts a powerful AGN and is highly
absorbed with an inferred column density of 
$N_H\sim 2\times 10^{24}$\ cm$^{-2}$ 
(\markcite{vignati99}Vignati et al.\ 1999).
On the figure we have connected with a straight line
the values of $L_{HX}$ calculated from the observed flux 
(\markcite{iwasawa99}Iwasawa 1999) and from the
inferred intrinsic flux
(\markcite{vignati99}Vignati et al.\ 1999) to illustrate
the large effects that absorption can make for highly obscured
AGN at low redshift.

Using the submillimeter inferred FIR luminosity of source 8 
and the radio inferred FIR luminosities of the other 
hard X-ray sources, we find that
the values of $L_{FIR}/L_{HX}$ for all the sources are comparable
to that of NGC~6240. Also shown in Fig.~\ref{fig15}a (open diamonds) 
are the luminous $z>4$ radio quiet quasars observed in the 
submillimeter by \markcite{mcmahon99}McMahon et al.\ (1999) 
for which there are also X-ray detections
(\markcite{kaspi00}Kaspi et al.\ 2000; we converted the
$f_\nu$ at 2\ keV values given in their Table~2 to $2-10$\ keV 
fluxes assuming $\Gamma=2$). 
These quasars are in the region of 3C~273 in Fig.~\ref{fig15}a.
The two gravitationally-lensed sources in the field of A370
(open triangles) detected in hard X-rays by 
\markcite{bautz00}Bautz et al.\ (2000) and in the submillimeter by 
\markcite{smail97}Smail, Ivison, \& Blain (1997)
have higher $L_{FIR}/L_{HX}$ ratios than the typical
X-ray source. However, none of the hard X-ray selected
sources remotely approaches the high 
$L_{FIR}/L_{HX}=3.4\times 10^4$ value of Arp~220.

Figure~\ref{fig15}b shows $L_{FIR}/L_{HX}$
versus $L_{HX}$ for the above submillimeter sources and for 
submillimeter sources in the literature that have
millimetric redshifts
(\markcite{bcr00}Barger, Cowie, \& Richards 2000 and 
\S~\ref{secsmmsample}) and X-ray detections or limits 
(\markcite{horn00}Hornschemeier et al.\ 2000 and 
\S~\ref{secsmmsample}). 
If most of the submillimeter sources are star-formers like Arp~220, 
then much deeper hard X-ray observations would be required to 
detect them.

\section{A Submillimeter Selected Sample: Hard X-ray Properties and
Contribution to the Hard XRB}
\label{secsmmsample}

\begin{figure*}[tb]
\centerline{\psfig{figure=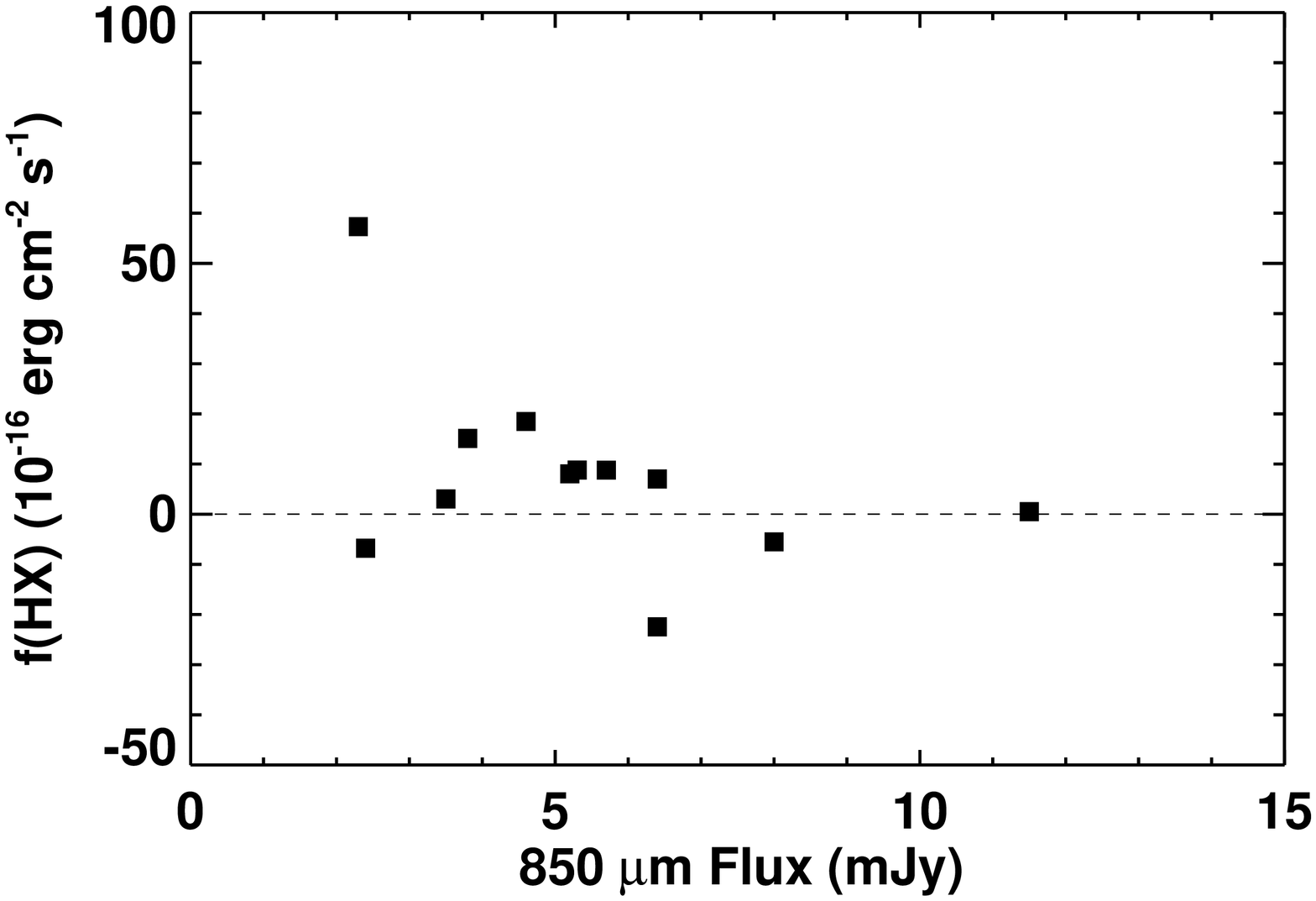,width=3.5in,angle=90}}
\figurenum{16}
\figcaption[]{Hard X-ray versus submillimeter fluxes
for the submillimeter selected sample in the SSA13 X-ray field.
The strongest hard X-ray emitter is source 8. The second
strongest hard X-ray emitter is not in Table~1 but is a known soft
X-ray emitter.
\label{fig16}
}
\end{figure*}

The foregoing section presented the submillimeter nature of
the hard X-ray sources. We now invert the approach and ask
what are the hard X-ray properties of a submillimeter 
selected sample.
We may address this by looking at the ensemble properties
of submillimeter selected sources in the SSA13 field. There
are twelve $3\sigma$ source detections at 850\ $\mu$m
in the S2 and S3 chips within a $4.5'$ radius of the optical axis. 
The fluxes range from just over 2.3\ mJy to 11.5\ mJy. 
The positional accuracies for the submillimeter sources are
relatively poor because of the large beam size. 
Six of the twelve submillimeter sources have 20\ cm
counterparts within $5''$, and the dispersion
of the offsets is $3.4''$ (see also
\markcite{bcr00}Barger, Cowie, \& Richards 2000).
In order to allow for the positional uncertainty, 
we determined the $2-10$\ keV fluxes
for each of the submillimeter sources in $10''$
diameter apertures. The results are shown in Fig.~\ref{fig16}.
The only submillimeter source that is also a strong hard X-ray 
emitter is source 8 from Table~1. The submillimeter source with
the second strongest hard X-ray flux in Fig.~\ref{fig16} is not in 
the hard X-ray sample but is a known soft X-ray emitter (source 28 
in Table~1 of MCBA).

Because of the large apertures used, there is a significant 
probability of random overlap with a non-coincident X-ray source. 
In order to quantify this, we have again run Monte Carlo simulations
using an equal number of sources at random positions. The
$1\sigma$ dispersion is 
$0.5 \times 10^{-16}$\ erg\ cm$^{-2}$\ s$^{-1}$\ mJy$^{-1}$.
Only 5\% of the simulations have values in excess of 
$1.2\times 10^{-16}$\ erg\ cm$^{-2}$\ s$^{-1}$\ mJy$^{-1}$. 
In fact, the two sources that are contributing all 
of the X-ray signal have 20\ cm radio counterparts that are closely 
coincident with the X-ray sources. Thus, both of these are real 
identifications rather than chance contamination.

The ratio of the total hard X-ray
flux in the sample to the total submillimeter flux is 
$1.4\pm 0.5\times 10^{-16}$\ erg\ cm$^{-2}$\ s$^{-1}$\ mJy$^{-1}$.
We may use this ratio to estimate the fraction
of the hard XRB that arises from submillimeter
sources in this flux range. The EBL of the $2-10$\ mJy
source population is $9.3\times 10^3$ mJy\ deg$^{-2}$
(\markcite{bcs99}Barger, Cowie, \& Sanders\ 1999),
which would contribute 
$1.3\times 10^{-12}$\ erg\ cm$^{-2}$\ s$^{-1}$\ deg$^{-2}$
in the $2-10$\ keV band or 6\% of the hard XRB.
This contribution would rise to 19\% to 27\% if we 
normalized to the total submillimeter background of
$3.1\times 10^4$\ mJy\ deg$^{-2}$ of
\markcite{puget96}Puget et al.\ (1996) or
$4.4\times 10^4$\ mJy\ deg$^{-2}$ of
\markcite{fixsen98}Fixsen et al.\ (1998).

However, nearly all of the X-ray signal from the submillimeter 
sample is coming from source 8.
If this single source is removed, the total hard X-ray
to total submillimeter flux ratio drops to
$0.6\pm 0.5\times 10^{-16}$\ erg\ cm$^{-2}$\ s$^{-1}$\ mJy$^{-1}$.
If we place the submillimeter sources at $z=2$ 
(consistent with the $z=1-3$ spectroscopic and millimetric redshifts 
from, e.g., \markcite{barger99}Barger et al.\ 1999 and
\markcite{bcr00}Barger, Cowie, \& Richards 2000),
then the $1\sigma$ lower limit on the
ratio of $L_{FIR}$ to $L_{HX}$ would be approximately 1100.
This lower limit is above the obscured AGN values
and approaching that of Arp~220. It therefore appears
that most of the submillimeter sources, at least above
2\ mJy, are star formers with a small admixture of
obscured AGN. Similar conclusions have recently been reached by
\markcite{fabian00}Fabian et al.\ (2000), 
\markcite{horn00}Hornschemeier et al.\ (2000), and
\markcite{sever00}Severgnini et al.\ (2000).

\section{A Radio Selected Sample: Hard X-ray Properties and
Contribution to the Hard XRB}

There are 107 20\ cm sources above the $5\sigma$ radio
limit of $25\ \mu$Jy that lie on the S2 and S3 chips 
within a $4.5'$ radius of the optical axis.
We have measured their $2-10$\ keV fluxes using $2.5''$ 
diameter apertures. Above a hard X-ray flux of 
$3\times10^{-15}$\ erg\ cm$^{-2}$\ s$^{-1}$,
we recover only the 10 hard X-ray sources
with $5\sigma$ radio counterparts in Table~1. Above
$1 \times10^{-15}$\ erg\ cm$^{-2}$\ s$^{-1}$, 16 sources
are detected. The contribution to the hard XRB
from the entire radio selected sample is 
$5.9\times 10^{-12}$\ erg\ cm$^{-2}$\ s$^{-1}$\ deg$^{-2}$
(26\%), nearly all of which comes from the 
hard X-ray sources in Table~1.
The small fraction ($10-15$\%) of radio sources that are hard 
X-ray sources, and therefore AGN, is consistent with expectations 
that the great bulk of the 20\ cm sources at these faint fluxes 
are due to synchrotron emission arising from supernova remnants 
in the interstellar medium of the galaxies rather than to nuclear 
activity. The radio morphologies, which will provide more insight 
into the nature of the hard X-ray sources, will be discussed in 
\markcite{richards01}Richards et al.\ (2001).

\section{Hard X-ray Contribution to the EBL}
\label{secsmmebl}

\begin{figure*}[tb]
\centerline{\psfig{figure=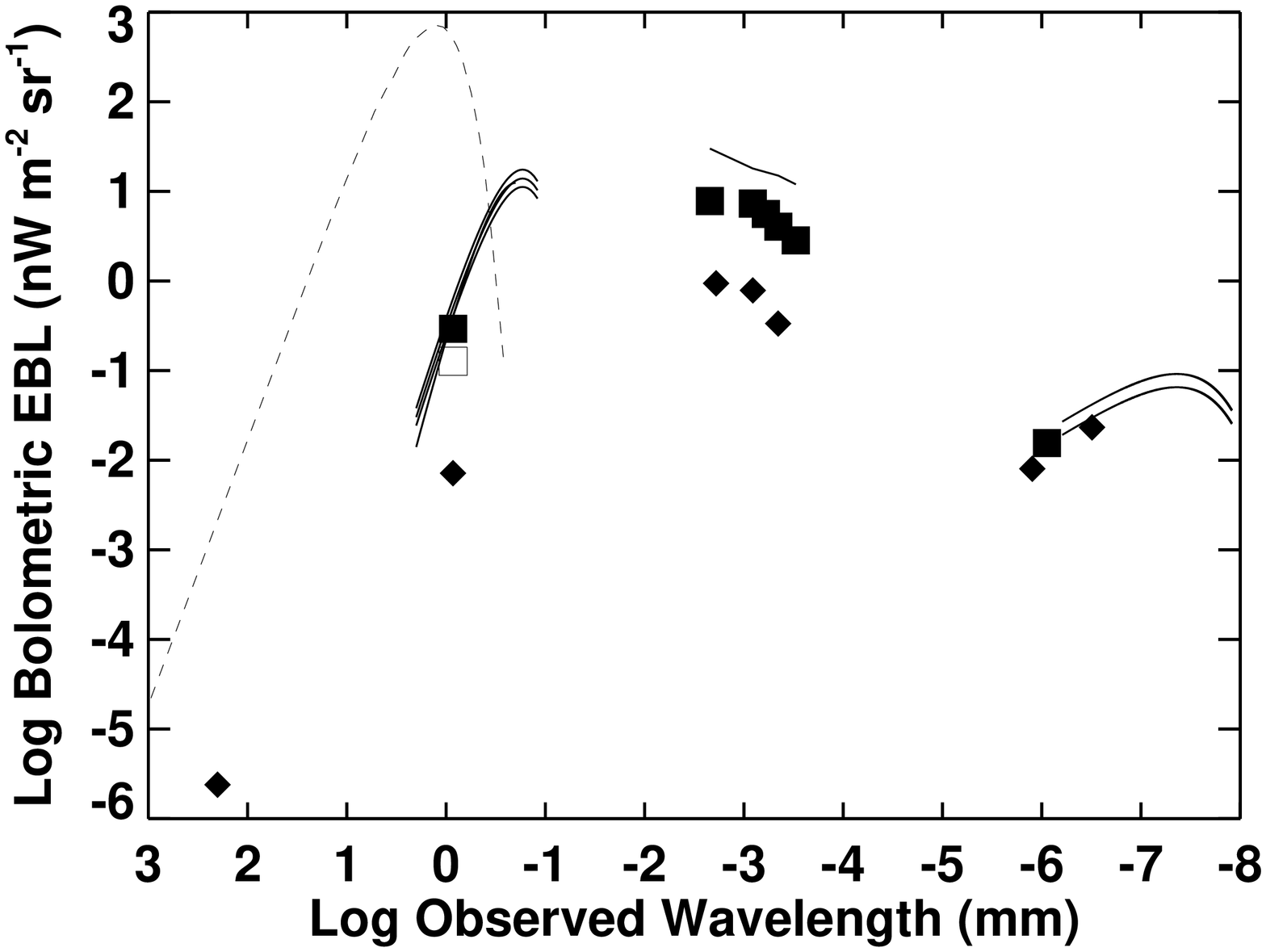,width=3.5in,angle=90}}
\figurenum{17}
\figcaption[]{Filled diamonds show the contribution of our
{\it Chandra} hard X-ray selected sample to the EBL at each of the
measured wavelengths. The error bars are smaller than the symbol
size at all wavelengths. The dashed line shows the cosmic
microwave background. Solid curves show the measured
total extragalactic light in each wavelength regime:
submillimeter from Puget et al.\ (1996) and
Fixsen et al.\ (1998); optical/NIR from
Bernstein et al.\ (1999) and Wright \& Reese\ (2000);
X-ray from Marshall et al.\ (1980), both
at the original flux and re-normalized to the
more recent higher flux determinations discussed in the text.
Filled squares show the integrated contribution of the source
counts: submillimeter counts to 0.5\ mJy from Blain et al.\ (1999),
but since these counts are quite uncertain, the well-established
$>2$\ mJy counts from Barger, Cowie, \& Sanders (1999) are also
shown ({\it open square});
optical/NIR from Williams et al.\ (1996) and Gardner et al.\ (1993);
$1-2$\ keV X-ray from Table~2 of MCBA.
\label{fig17}
}
\end{figure*}

The summed contributions of our {\it Chandra} hard X-ray sources 
to the EBL are shown versus wavelength in
Fig.~\ref{fig17} as the filled diamonds. These are compared with
measurements of the EBL (solid curves) and with the
integrated light from direct counts (large squares); the latter
now lie close to the EBL at all wavelengths.

For galaxies at $z\gg 1$ the X-ray power reprocessed by dust 
is expected to appear at submillimeter wavelengths.
We computed the hard X-ray contribution
to the submillimeter EBL by considering the average
submillimeter properties of the whole X-ray sample and
of selected subsamples. Error-weighted sums of
the submillimeter sources are given in Table~\ref{tab3}, together
with the correspondingly weighted total hard X-ray fluxes 
for the sample and the ratios of the two. The total hard
X-ray sample has a significant ($>3\sigma$) submillimeter flux
of 20\ mJy. As expected, most
of this arises in the $z>1.5$ sources and in the unidentified
sources. The total 850\ $\mu$m flux of the sample
corresponds to an EBL of $4\times 10^{3}$\ mJy\ deg$^{-2}$,
or from 9 to 13\% of the 850\ $\mu$m EBL, depending on whether 
the normalization of \markcite{fixsen98}Fixsen et al.\ (1998)
or \markcite{puget96}Puget et al.\ (1996) is adopted.
These values are similar to the obscured AGN model
predictions of \markcite{almaini99}Almaini et al.\ (1999)
and \markcite{gunn00}Gunn \& Shanks (2000).
However, it appears the the great bulk of the 850\ $\mu$m
EBL must arise in sources which are too faint at
$2-10$\ keV to be seen at the current submillimeter threshold.

We computed the hard X-ray contribution to the optical/NIR EBL 
by summing the fluxes corresponding to the corrected $3''$ diameter 
aperture magnitudes at these wavelengths.
This light is dominated by the brighter galaxies, many of
them with `normal' galaxy spectra, so there is almost
certainly strong contamination of the estimate of the AGN
contribution by the light of the host galaxies. 
The optical/NIR EBL should therefore be considered
strictly as an upper bound.

In summary, the hard X-ray sample
contributes about 10\% of the light at both UV/optical 
and submillimeter wavelengths. Both wavelength regimes have 
similar total bolometric light densities.
However, because there is galactic light contamination in
the optical/NIR, it is likely that the FIR
is where most of the AGN light is emerging from the hard X-ray 
sources.

\begin{deluxetable}{ccccc}
\renewcommand\baselinestretch{1.0}
\tablewidth{0pt}
\parskip=0.2cm
\tablenum{3}
\small
\tablehead{
\colhead{Sample} & Total & Total &
$\sum f(2-10$\ keV)/ \cr
& $f(850\ \mu$m) & $f(2-10$\ keV) & $\sum f(850\ \mu$m) & \cr
& (mJy) & ($10^{-16}$\ ergs\ cm$^{-2}$\ s$^{-1}$) &
($10^{-14}$\ ergs\ cm$^{-2}$\ s$^{-1}$\ mJy$^{-1}$) &
}
\startdata
All & $20\pm 6$ & 1800 & $0.9^{1.3}_{0.7}$  \cr
$z<1.5$ & $1\pm 6$ & 840 & $6.0^{\infty}_{1.2}$ \cr
$z>1.5$\ (non-QSOs) & $7\pm 3$ & 340 & $0.5^{0.8}_{0.3}$ \cr
QSOs & $5\pm 2$ & 290 & $0.6^{1.0}_{0.4}$ \cr
Unidentified & $8\pm 3$ & 600 & $0.8^{1.2}_{0.5}$ \cr
\enddata
\label{tab3}
\end{deluxetable}

\section{Black Hole Mass Accretion}

\begin{figure*}[tb]
\centerline{\psfig{figure=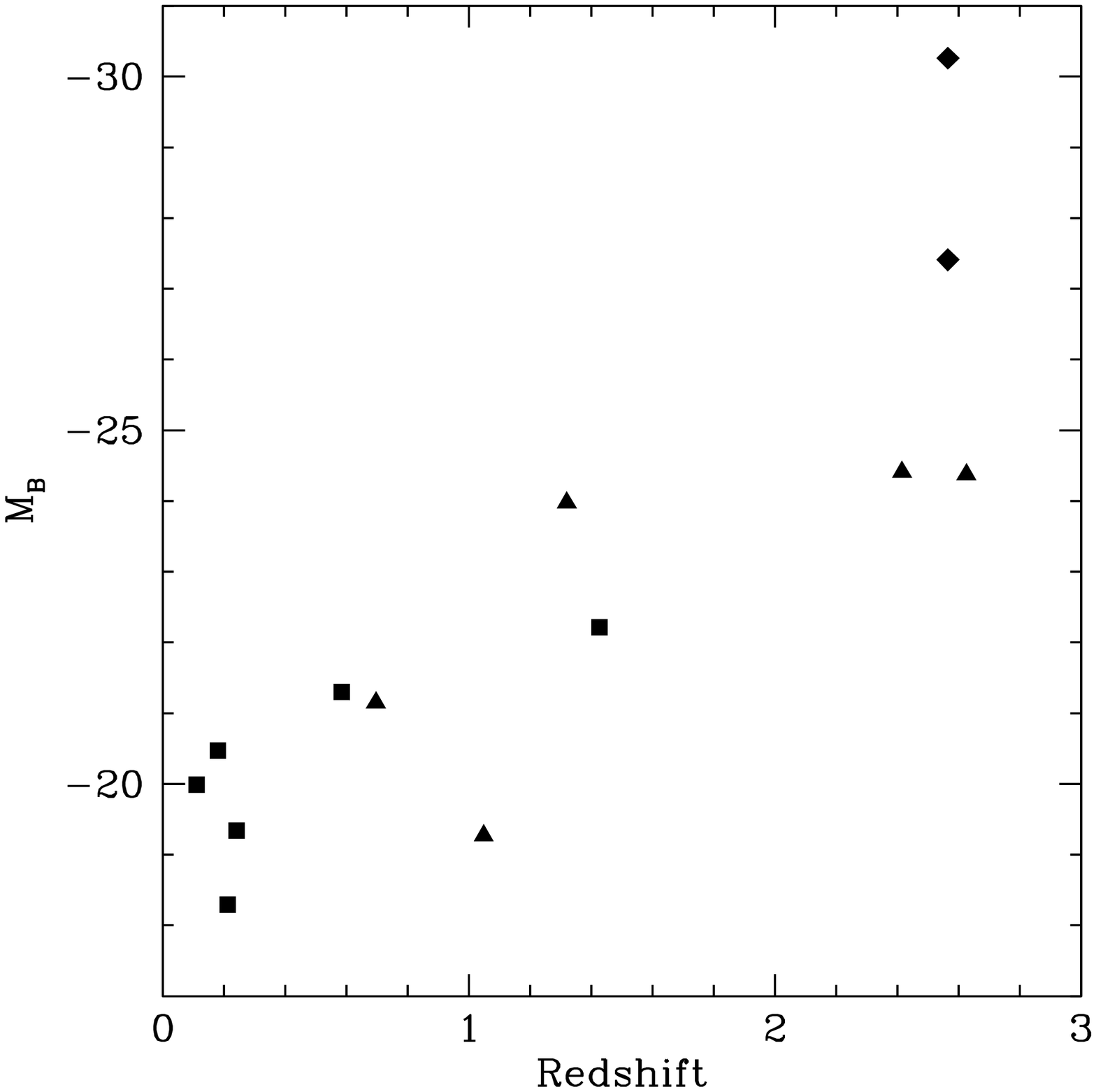,width=3in,height=3in}}
\figurenum{18}
\figcaption[]{
Absolute $B$-band magnitudes versus redshift.
`Normal' galaxies are denoted by filled squares,
AGN by filled triangles, and quasars by filled diamonds.
The K-corrections have been computed using the Coleman, Wu, \&
Weedman 1980 SED for an early spiral galaxy.
\label{fig18}
}
\end{figure*}

\begin{figure*}[tb]
\centerline{\psfig{figure=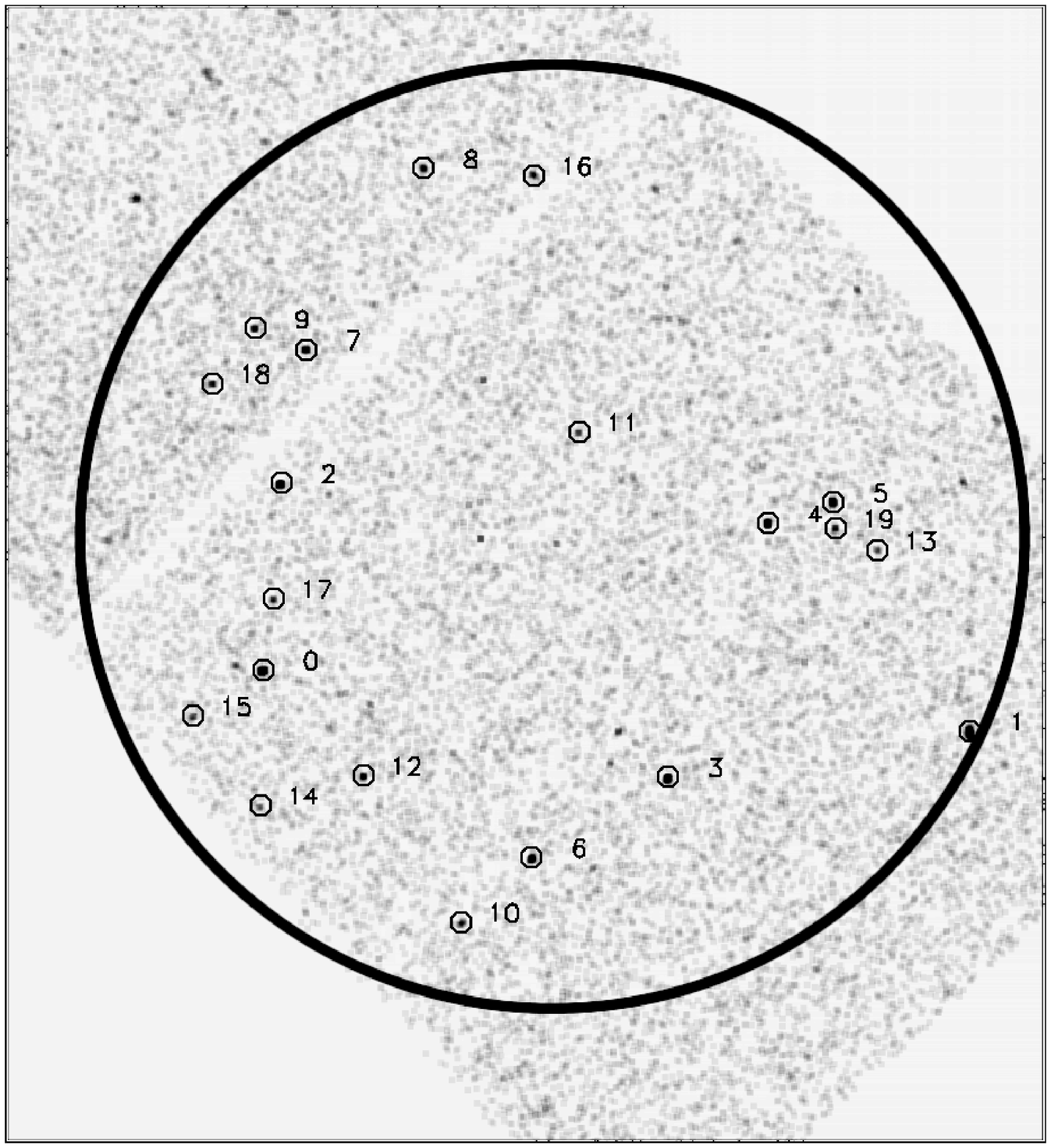,angle=90}}
\figurenum{1}
\figcaption[]{
{\it Chandra} hard X-ray image of the SSA13 field.
The 20 significant hard X-ray source positions are identified by
the small circles. The large circle illustrates the $4.5'$ radius
adopted for the analysis.
\label{fig1}
}
\end{figure*}

\begin{figure*}[tb]
\centerline{\psfig{figure=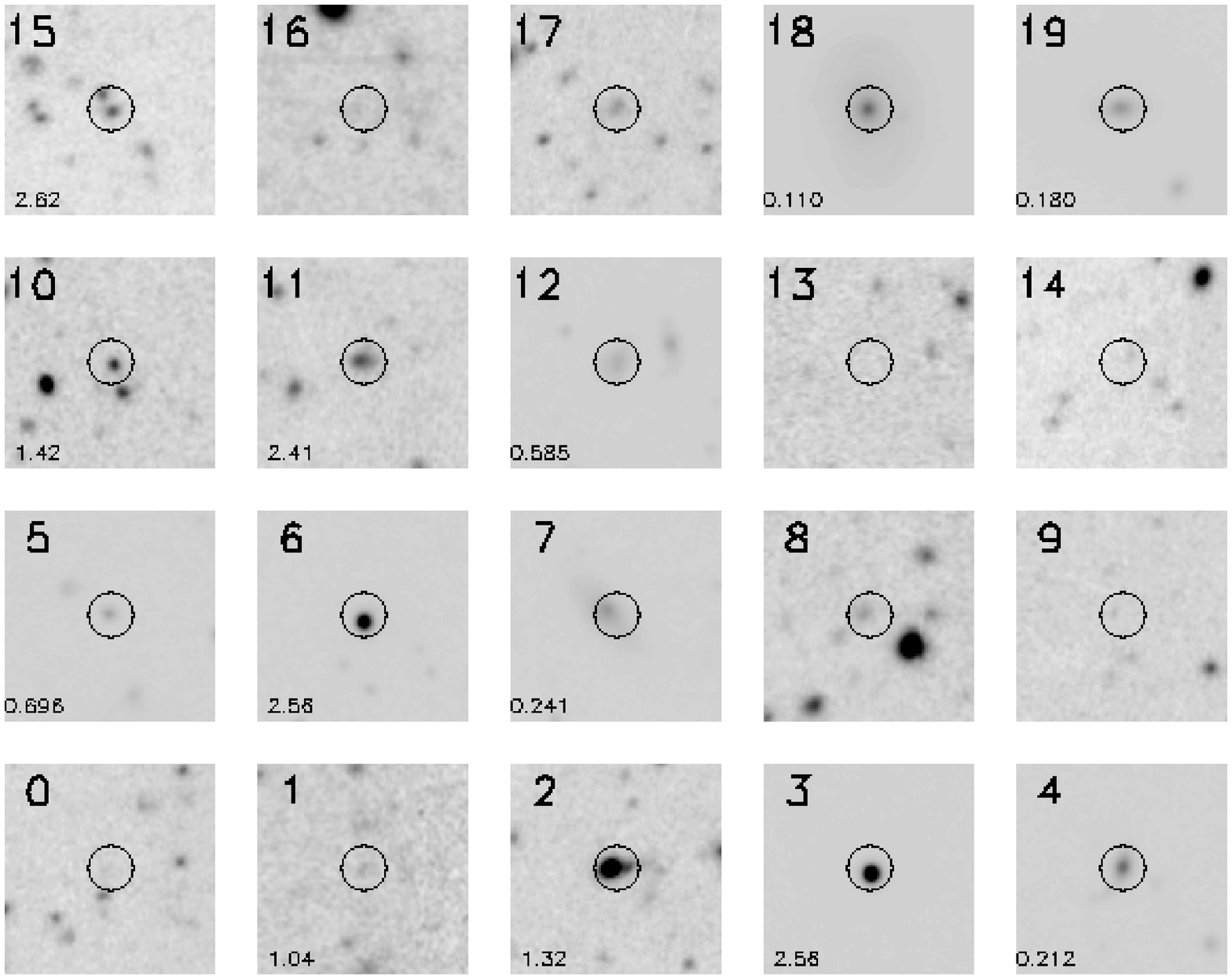,angle=90,width=7.5in}}
\figurenum{2}
\figcaption[]{
$9''\times 9''$ $B$-band thumbnail
images of the hard X-ray sources in Table~1.
The images are from ultradeep data obtained with LRIS on Keck.
The identification numbers are as in Table~1,
and the sources are ordered from lower left to upper right by
decreasing $2-10$\ keV flux. A circle of $1.5''$ radius, typical
of the maximum positional uncertainty, is superimposed on
each thumbnail. Redshifts, where available, are given in the lower
left-hand corners of the thumbnails. North is up and East is to the left.
For the more luminous objects we have shown the images at higher
surface brightness to allow the positions of the objects with
respect to the centers of the galaxies to be seen.
\label{fig2}
}
\end{figure*}

\begin{figure*}[tb]
\centerline{\psfig{figure=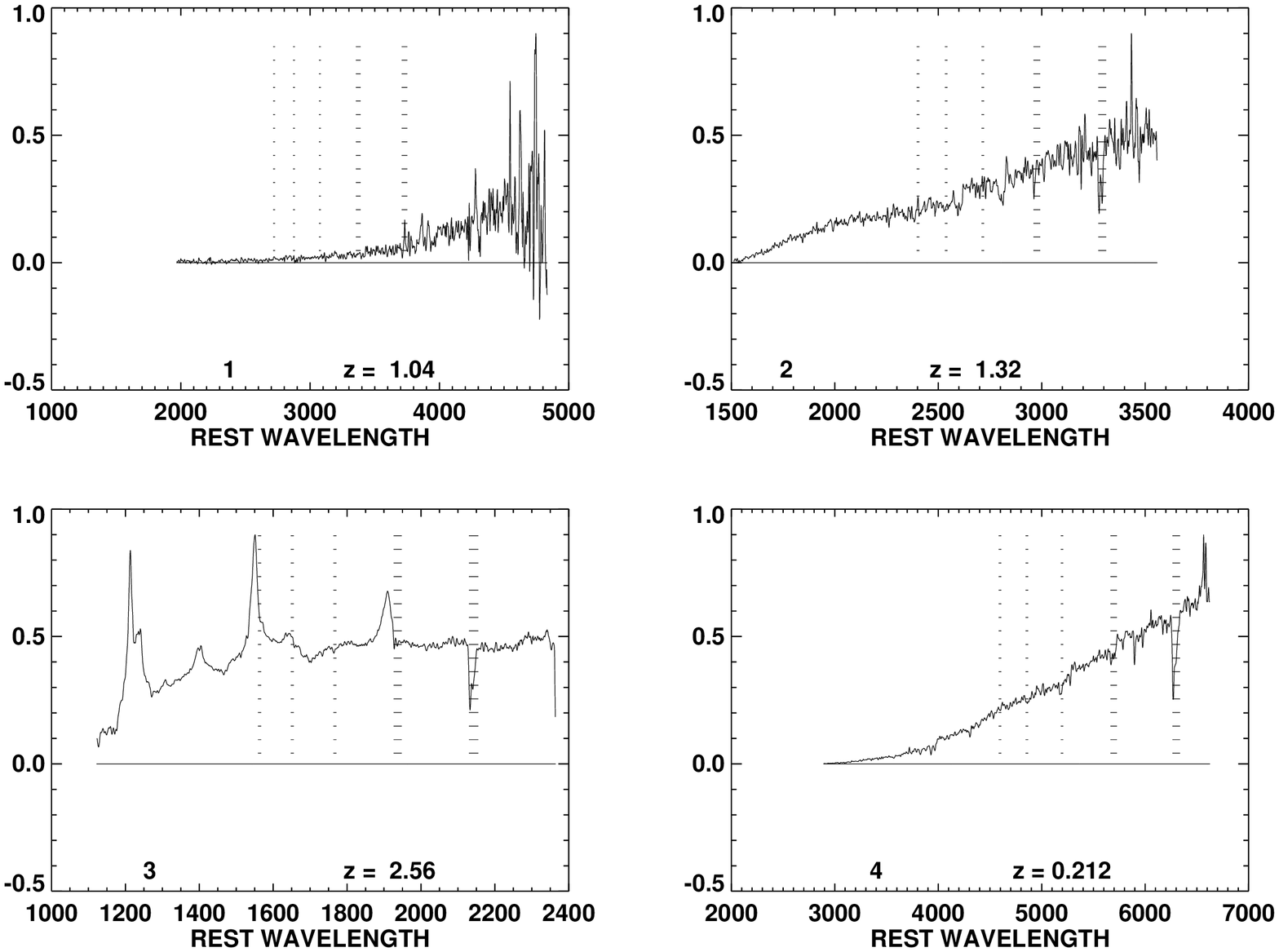,angle=90,width=7in}}
\figurenum{3}
\figcaption[]{
The Keck LRIS spectra of the 13 hard X-ray sources with secure
redshift identifications. The spectra are shown in the rest-frame
with source number and the measured redshift at the bottom of
the frame. Dashed lines mark strong atmospheric absorption 
or emission features.
\label{fig3}
}
\end{figure*}

For nearby normal galaxies, there is an approximate
empirical relation between the black hole mass, M$_{bh}$,
and the absolute magnitude of the bulge
component of the host galaxy, $M$(bulge)
(e.g., \markcite{magorrian98}Magorrian et al.\ 1998;
\markcite{wandel99}Wandel 1999;
\markcite{ferr00}Ferrarese \& Merritt\ 2000;
\markcite{gebhardt00}Gebhardt et al.\ 2000).
Assuming that the relation also holds at high redshift, we can 
estimate the black hole masses of our `normal' galaxies, subject 
to the uncertainties in translating our observed $B$-band magnitudes 
into $M_B$(bulge) that would worsen the empirical relation
(\markcite{kormendy00}Kormendy \& Ho 2000).
The $M_{B}$ values for our hard X-ray sample are plotted versus
redshift in Fig.~\ref{fig18}.
The inferred ${\rm M}_{bh}$ values in our `normal' galaxies 
are in the approximate range $5\times 10^{7}$\ M$_\odot$ to
$10^{9}$\ M$_\odot$.

As discussed in \S~\ref{secoptsample}, luminous hard X-ray sources
are common in bulge-dominated optically luminous galaxies
with about 10\% of the population showing activity
at any given time. This preferential activity of
luminous sources with massive bulges
presumably reflects the relation
between the central massive black hole and the luminosity
of the bulge. If the fraction of galaxies showing such behavior
reflects the fraction of time that each galaxy spends
accreting onto its massive black hole, then
we require each such galaxy to be active for somewhere
between 1 and 2\ Gyr. This is considerably longer than the
theoretically estimated accretion time of 0.01\ Gyr for
black hole fuelling by mergers
(\markcite{kauff00}Kauffmann \& Haenelt 2000)
and therefore may suggest that the activity is being powered by
smaller mergers or by internal flows within the galaxies.

We may address the issue of the total accreted mass required to
account for the luminosity in each galaxy by conservatively assuming
that the hard X-ray flux-weighted average of $L_{FIR}/L_{HX}=33$ derived
in \S~\ref{secbolo} is the ratio of the AGN's bolometric luminosity
to its $2-10$\ keV luminosity; hereafter, we denote this 
bolometric correction by $A$. 
The mass inflow rate is
$\epsilon\Delta M_{bh} c^2/\Delta t=L_{BOL}=A L_{HX}$,
where $\epsilon$ is the efficiency for re-radiation of the
accretion energy. The total mass flow over the accretion time
$\Delta t\sim 1.5$\ Gyr is

$$\Delta M_{bh}=10^7\ {\rm M}_\odot \Biggl({L_{HX} \over {10^{42}\ 
{\rm erg\ s^{-1}}}}\Biggr)\Biggl({A \over 33}\Biggr)\Biggl({{\Delta t} 
\over {1.5\ {\rm Gyr}}} \Biggr)\Biggl({0.1 \over \epsilon}\Biggr)$$

\noindent
Our `normal' galaxies have hard X-ray luminosities ranging from
$0.1$ to $50\times 10^{42}\ {\rm erg\ s^{-1}}$, 
for which the above accretion
masses are $10^6$ to $5\times 10^8\ M_\odot$.
Thus, even for the maximum plausible efficiency, $\epsilon\sim 0.1$, we
are seeing a substantial 
fraction of the growth of these supermassive black holes
(\markcite{fabian99}Fabian \& Iwasawa\ 1999).

We may quantify this in a fairly model-independent way as follows. 
The bolometric surface brightness at the
present time is simply related to the $2-10$\ keV XRB by

$$I_{BOL}=0.85\ A\ I_{HX}$$

\noindent
where the factor 0.85 is discussed in \S~\ref{seclum}.
The bolometric EBL is related to the energy production as

$$I_{BOL}={{\epsilon\ c^3\ \rho_{bh}} \over {4\ \pi\ (1+\bar{z})}}$$

\noindent
where $\rho_{bh}$ is the universal mass density of supermassive black
holes, and $\bar{z}$ is the mean redshift of the contributing 
sources. The $(1+\bar{z})$ factor reflects the adiabatic expansion 
loss. The hard X-ray flux-weighted mean
redshift is $\bar{z}=1.3$ for the spectroscopically
identified sources (including source 8) and
$\bar{z}=1.5$ if we place the remaining sources at $z=2$.

We combine the two equations and normalize to the
observed $2-10$\ keV background from 
\markcite{vecchi99}Vecchi et al.\ (1999),
$I_{HX}=7.6\times 10^{-8}$\ erg\ cm$^{-2}$\ s$^{-1}$\ sr$^{-1}$,
to find

$$\rho_{bh} = 2\times 10^{-35}\ \Biggl({{1+\bar{z}}\over 2.5}\Biggr)\ 
\Biggl({0.1 \over \epsilon}\Biggr)\ {\rm g\ cm}^{-3}$$

\noindent
where we have again conservatively used $A=33$.
Inclusion of the bolometric optical light would 
raise this estimate by about 40\%.

If we adopt a spheroid mass density of $10^{-32}$\ g\ cm$^{-3}$,
which has a plausible multiplicative error of about 2
(\markcite{cowie88}Cowie 1988; \markcite{fuk98}Fukugita et al.\ 1998),
the black hole-bulge relation expressed in mass terms

$${\rm M}_{bh} = 0.002-0.006\ {\rm M(bulge)}$$

\noindent
gives a black hole mass density of
$\rho_{bh} = 2-6\times 10^{-35}$\ g\ cm$^{-3}$.
Thus, even
for high-end estimates of the radiative efficiency ($\epsilon\sim 0.1$)
we are seeing a significant fraction of the black hole accretion.
Conversely, since we are probably missing some fraction
of the activity from sources that are too obscured
to be seen in the $2-10$\ keV sample, we must have a
high radiative accretion efficiency ($\epsilon\gg 0.01$) or the 
required black hole mass density would be too high.

We may go one step further and ask when the black
holes formed, since the above equations also apply
to the sources divided into redshift slices. We know
directly from the spectroscopic observations that
at least 16\% of the $2-10$\ keV background arises
below $z=1$, 18\% between $z=1$ and 2, and
13\% between $z=2$ and 3. The remaining 21\% of the
$2-10$\ keV light seen in the current sample
most plausibly lies in the $z=1.5-3$ range.
Weighting by the mean redshifts in each interval,
we find that at least 10\% of the observed black hole mass
formation occurs at $z<1$. This fraction could
rise if there are more obscured hard X-ray sources
missing from the $2-10$\ keV sample which
preferentially lie at low redshift.

\noindent

\section{Conclusions}

We carried out an extensive multi-wavelength observational program 
to determine the nature of the hard X-ray background sources.
Our major conclusions are as follows

$\bullet$ In the 57\ arcmin$^2$ {\it Chandra} SSA13 field, 
we detected 20 sources with $2-10$\ keV fluxes greater than
$3.8\times 10^{-15}$\ erg\ cm$^{-2}$\ s$^{-1}$.

$\bullet$ We spectroscopically identified the 13 hard X-ray 
sources brighter than $I=23.5$\ mag; all are in the redshift 
range $z=0-3$. The spectra fall
into three general categories: (i)\, 2 quasars, (ii)\, 5 AGN,
and (iii)\, 6 optically `normal' galaxies. 
For the latter category, the AGN are
either very weak or undetectable in the optical.

$\bullet$ The soft to hard X-ray flux ratios of the hard sample 
can be described by a rather
narrow range of neutral hydrogen column densities from
$N_H=2\times 10^{22}$\ cm$^{-2}$ to $3\times 10^{23}$\ cm$^{-2}$.
At most three of the 14 sources at $z>1$ have column
densities above $N_H=3\times 10^{23}$\ cm$^{-2}$, which suggests
that we are seeing most of the obscured AGN in the present sample.

$\bullet$ The hard X-ray sample is consistent with a constant 
surface density of $\sim 400$ sources per square degree per unit 
redshift. The redshift distribution is very similar to 
that of previous soft selected samples with similar limiting
sensitivities.

$\bullet$ The hard X-ray luminosities of the spectroscopically
identified sources range from just over $10^{41}$\ erg\ s$^{-1}$
to $\sim 10^{45}$\ erg\ s$^{-1}$. All but two
are more X-ray luminous than the local galaxy populations.

$\bullet$ The integrated $2-10$\ keV light of the 20 hard
X-ray sources corresponds to an EBL of 
$1.34\times 10^{-11}$\ erg\ cm$^{-2}$\ s$^{-1}$\ deg$^{-2}$,
or between 58 to 84 percent of the hard XRB, depending on the
XRB normalization. Most of this light is from the
$z<2$ population.

$\bullet$ The colors of the spectroscopically unidentified
sources suggest that they are likely early galaxies at $z=1.5-3$.

$\bullet$ The `normal' galaxies and AGN follow the upper
envelope of the star forming field galaxy population.

$\bullet$ Between 4 and 12 percent of $I<23$ galaxies 
with $-22.5>M_I>-24$ have hard X-ray fluxes above
$1.0\times 10^{-15}$\ erg\ cm$^{-2}$\ s$^{-1}$.
Nearly all of these lie in the redshift range
$z=0.3-1.5$, for which the flux threshold
corresponds to $L_{HX}\sim 0.3-3\times 10^{42}$\ erg\ s$^{-1}$.

$\bullet$ Excluding the two quasars, the hard X-ray flux-weighted
average bolometric optical to $2-10$\ keV luminosity ratio for
the hard X-ray sources is low (9), providing strong evidence that
the sources are powered by AGN.

$\bullet$ Excluding the two quasars, the
hard X-ray flux-weighted average bolometric correction from the
$2-10$\ keV luminosity to the bolometric light of the AGN is
somewhere between 33 and 42, depending on the optical light
contamination by stars.

$\bullet$ Of the 20 hard X-ray sources in our sample, only one
(source 8) was significantly detected in the submillimeter.
The millimetric redshift of source 8, obtained from the
submillimeter to radio flux ratio, is in the range $1.2-2.4$.
Its rest-frame SED is similar to that of the heavily
dust obscured local AGN NGC~6240.

$\bullet$ Bolometric FIR luminosities or upper limits inferred 
from the radio data using the FIR-radio correlation or
from the submillimeter data using the NGC~6240 
or PG~1543+489 SEDs and luminosities are consistent.

$\bullet$ The $>2$\ mJy submillimeter source population contributes 
very little to the hard XRB (6\%).

$\bullet$ Excluding the one submillimeter source
that had a significant hard X-ray detection,
the FIR to hard X-ray luminosity ratio 
for the submillimeter selected sample has a
$1\sigma$ lower limit of 1100; this is above the obscured
AGN values and approaching that of Arp~220. Thus, it appears
likely that most of the $>2$\ mJy submillimeter sources are 
star formers.

$\bullet$ The radio selected sample contributes 26\%
to the hard XRB, nearly all of which comes from the observed 
hard X-ray sources.

$\bullet$ The hard X-ray sources contribute about 10\% of 
the light at both UV/optical and submillimeter wavelengths.
However, because there is stellar light contamination in
the optical/NIR, it is likely that most of the AGN light is
emerging in the FIR.

$\bullet$ The masses of the black holes in our `normal' galaxies
are estimated from the Magorrian relation to be in the range
$5\times 10^{7}$\ M$_\odot$ to $10^{9}$\ M$_\odot$.

$\bullet$ Luminous hard X-ray sources are common in bulge dominated
optically luminous galaxies with about 10\% of the population
showing activity at any given time. Since
the hard X-ray emission is likely associated with episodes of
accretion onto the central massive black holes in these galaxies,
the 10\% may represent the duty cycle
of galaxies that are active at any given time.
Even with a high-end estimate of the radiative efficiency
($\epsilon=0.1$), the black hole mass density required to account
for the observed light is comparable to the local black hole
mass density.

\acknowledgements
We thank Keith Arnaud for generating the X-ray images for 
this analysis, and we thank Nicolas Biver and James Deane for 
taking the SCUBA observations. We thank
an anonymous referee for helpful comments about the manuscript.
AJB and EAR acknowledge support from NASA through Hubble
Fellowship grants HF-01117.01-A and HF-01123.01-A awarded by the
Space Telescope Science Institute, which is operated by the
Association of Universities for Research in Astronomy, Inc.,
for NASA under contract NAS 5-26555.
AJB and LLC acknowledge support from NSF through grants
AST-0084847 and AST-0084816. The JCMT is operated by the Joint
Astronomy Center on behalf of the UK Particle Physics and
Astronomy Research Council, the Netherlands Organization for
Scientific Research, and the Canadian National Research Council.

\newpage

\newpage

\newpage

\begin{figure*}[tb]
\centerline{\psfig{figure=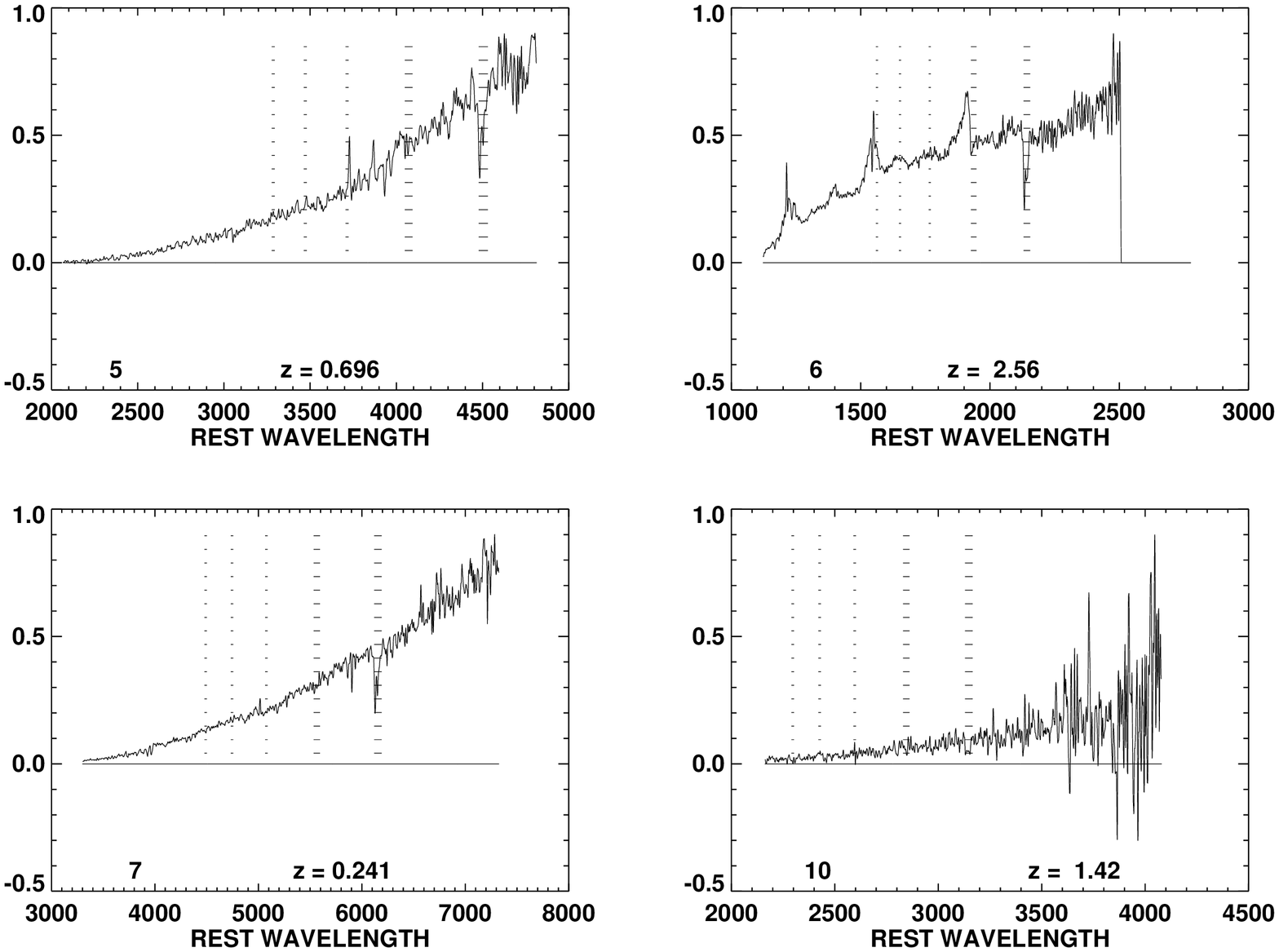,angle=90,width=7in}}
\figurenum{3}
\figcaption[]{
}
\end{figure*}

\newpage

\begin{figure*}[tb]
\centerline{\psfig{figure=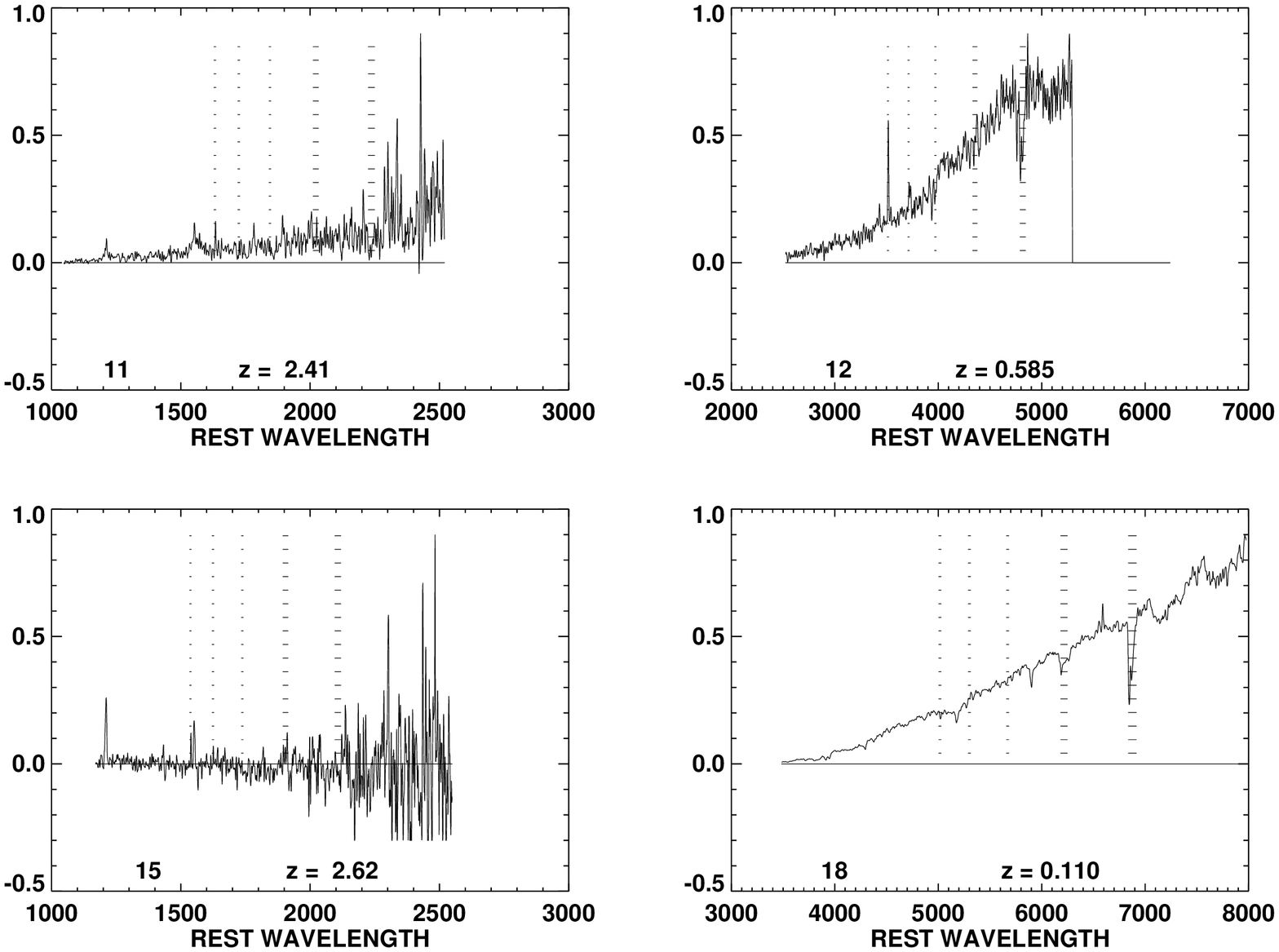,angle=90,width=7in}}
\figurenum{3}
\figcaption[]{
}
\end{figure*}

\begin{figure*}[tb]
\centerline{\psfig{figure=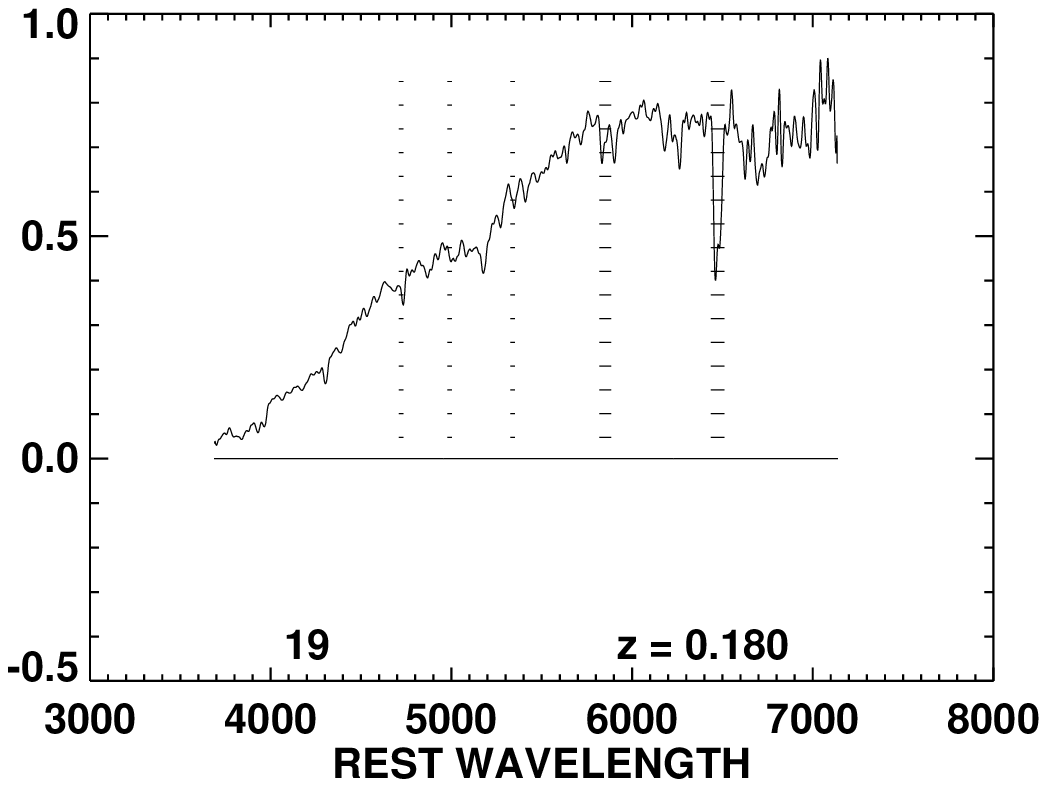,angle=90,width=7in}}
\figurenum{3}
\figcaption[]{
}
\end{figure*}

\end{document}